\documentclass[onecolumn]{article}    

\usepackage{graphicx}          
                               
\usepackage{amssymb}
\usepackage{amsthm}
\usepackage{gensymb}
\usepackage{subfig}
\usepackage{mathtools}
\usepackage{authblk}
\usepackage{hyperref}

\usepackage[a4paper, rmargin = 2.5cm, lmargin = 2.5cm]{geometry}

\theoremstyle{definition}

\theoremstyle{proposition}

\theoremstyle{assumption}

\theoremstyle{remark}

\begin{document}


\title{Autonomous Rapid Exploration in Close-Proximity of an Asteroid}  

\author[ a]{Rodolfo Batista Negri\footnote{email: rodolfo.negri@unifesp.br. \\ Personal website: \url{rodolfobnegri.com}}}
\author[ b]{Antônio F. B. de A. Prado}
\author[ b]{Ronan A. J. Chagas}
\author[ a]{Rodolpho V. de Moraes}
\affil[ a]{Federal University of São Paulo, ICT, 12247-014, São José dos Campos, São Paulo, Brazil}
\affil[ b]{National Institute for Space Research, 1758, 12227-010, São José dos Campos, SP, Brazil}
\date{}


      
\maketitle

\begin{abstract}                          
The increasing number of space missions may overwhelm ground support infrastructure, prompting the need for autonomous deep-space guidance, navigation, and control (GN\&C) systems. These systems offer sustainable and cost-effective solutions, particularly for asteroid missions that deal with uncertain environments. This study proposes a paradigm shift from the proposals currently found in the literature for autonomous asteroid exploration, which inherit the conservative architecture from the ground-in-the-loop approach that relies heavily on reducing uncertainties before close-proximity operations. Instead, it advocates for robust guidance and control to handle uncertainties directly, without extensive navigation campaigns. From a series of conservative assumptions, we demonstrate the feasibility of this autonomous GN\&C for robotic spacecraft by using existing technology. It is shown that a bolder operational approach enables autonomous spacecraft to significantly reduce exploration time by weeks or months. This paradigm shift holds great potential for reducing costs and saving time in autonomous missions of the future.
\end{abstract}


\section{Introduction}





There has been a significant increase in the number of space missions proposed and planned to explore asteroids in recent decades. The growing interest stems from various reasons, such as studying the origins of life and water on Earth, gaining a better understanding of asteroid properties, and protecting the Earth from potential impacts. NASA has recently launched the DART spacecraft, which is the first space experiment focused on deflecting asteroids. DART will assess a kinetic impact deflection strategy, considered the most technologically ready approach. This strategy heavily relies on the properties of the specific asteroid in question~\cite{feldhacker2017shape,cheng2018aida} (shape, structure, spin state, among others). In the event of an actual asteroid threat, it would be advantageous to have a reconnaissance spacecraft before the kinetic deflection. Astronomical observations alone are often insufficient to accurately determine a small body's properties. As an example, we can look at the comet 67P-Churyumov-Gerasimenko, where there was a notable contrast between the observed shape and its actual shape when the Rosetta spacecraft arrived. The study conducted by Feldhacker et al.~\cite{feldhacker2017shape} shows the significance of asteroid's shape in the context of kinetic deflection. This serves as a compelling example that highlights the necessity of in-situ observation in certain situations.

In a hypothetical situation where a deflection is urgently required with limited warning time, an exploratory spacecraft may need to arrive days or weeks before the kinetic impact. Its purpose would be to gather as much information as possible about the asteroid's environment and properties, thus increasing the chances of a successful impact. In this reconnaissance mission profile, each day and hour gained becomes crucial in capturing high-quality images of the asteroid, deploying probes on its surface, or any other necessary in-situ activity. The success of deflection missions is not the only benefit of rapidly exploring an asteroid. Missions aimed at exploring multiple asteroids in densely populated regions with similar orbits (such as Trojan asteroids) can also gain advantages from multiple rendezvous and the collection of valuable scientific data before moving on to the next target. This approach enables the timely exploration of a greater number of targets by swiftly acquiring high-quality scientific information. 

In addition to these benefits, and more importantly, an autonomous and rapid approach to exploration can shape current scientific asteroid missions to be more cost-effective and time-efficient. Current missions have a conservative and cautious operational profile, often taking months of surveying and slowly approaching the target to constrain the uncertainties to very low levels before the primary goal of the mission~\cite{takei2020hayabusa2,williams2018osiris}. For instance, the OSIRIS-Rex mission took about four months to approach and make a preliminary survey of the asteroid Bennu before being inserted into its first orbit. The preliminary survey had approximately 20 days, in which the spacecraft made multiple flybys at a distance of roughly 7 km to reduce the uncertainty in the asteroid's mass to 2\% before a safe insertion into orbit~\cite{williams2018osiris}. 

This conservative approach is indeed justified. Current mission profiles rely on a ground-in-the-loop for the spacecraft operation. The data are downlinked in telemetry to the Earth, meticulously analyzed by the ground team, and sent back as telecommands to be executed by the spacecraft, as in any other space mission. These round-trips may experience a delay of up to 40 minutes for NEAs (near-Earth asteroids), which places a high operational limitation. On top of that, we have the particularities of small bodies. They are known for exhibiting a highly perturbed environment with significant uncertainties before the spacecraft rendezvous. For instance, uncertainties in the asteroid's mass may reach up to 2,000\% from current astronomical observations~\cite{board2019finding}. Autonomy in asteroid missions can mean a paradigm shift in this conservative and cautious scenario.

Full autonomy in asteroid missions has gained particular attention in the last few years. Scheres \& McMahon \cite{scheeres2019autonomous} assess different mission architectures for enabling autonomous exploration of small bodies, showing that a spacecraft can orbit and deploy objects on the surface of a small body with minimal instrumentation. Panicucci \cite{panicucci2021autonomous} analyses the use of a SLAM (simultaneous localization and mapping) technique so that an autonomous spacecraft map the environment while also localizing itself. Takahashi \& Scheeres \cite{takahashi2021autonomous} show that the autonomous exploration of a small NEA is possible within current technology limitations. Other works are more focused on proposing control and guidance laws that can support autonomous operation, with particular attention to robust control laws, due to the highly perturbed and uncertain environment~\cite{furfaro2013asteroid,furfaro2015hovering,gui2017control,zhang2019spacecraft,negri2021autonomous}. 

Although these are essential advancements, no attempt has been made yet to take full advantage of an autonomous mission while considering the navigational difficulties encountered in a small-body scenario. For instance, most works concerned with the control and guidance laws make minor considerations about the navigational capabilities around an asteroid. In other cases \cite{takahashi2021autonomous,takahashiinproceedings}, the autonomous mission profile is still highly attached to current ground-based approaches, with the spacecraft spending months characterizing the asteroid before insertion in a stable orbit. 

For instance, in the study conducted by Takahashi and Scheeres \cite{takahashi2021autonomous}, the spacecraft spent approximately five months in its close-approach phase transitioning between hovering stations after departing from a distance of 10 kilometers from the asteroid. During the initial part of this approach, which lasted over 40 days, the spacecraft transited in four hovering locations with the primary objective of significantly reducing the uncertainty in the asteroid's gravitational parameter. It is important to note that before this close-approach phase, referred to as ``close-hovering'' by the authors, the spacecraft spent one month in the far-approach stage, traveling from thousands of kilometers away to reach its first hovering station, during which the uncertainty in the spacecraft's state was already reasonably constrained~\footnote{The precise quantity and order of magnitude is difficult to ascertain from the available figures.}.

Departing from that conventional architecture, which relies on cautiously constraining uncertainties to a very low level, this study adopts robust guidance and control to examine the feasibility of rapidly approaching and orbiting an asteroid without prior knowledge of its properties, while taking into account the navigation challenges presented in such an environment. The spacecraft uses onboard batch-sequential filtering to estimate the asteroid's properties while rapidly approaching it from hundreds of kilometers of distance. The end of the mission consists of the spacecraft orbiting the asteroid in tight orbits, using a robust orbital station-keeping control \cite{negri2021autonomous}. 

To accomplish this mission profile, we assume that the spacecraft is equipped with a LiDAR (Light Detection and Ranging), two optical cameras, and an IMU. Their uncertainties are conservatively considered and modeled. The main idea is to show that a paradigm shift in asteroid missions is possible within current technology, abandoning the idea of constraining the uncertainties to a minimal level before close-proximity operation. This study will be successful if we can show that a spacecraft can go straight and rapidly to close-proximity operation, dispensing all the lengthy and conservative steps today adopted and still considered in autonomous studies to reduce uncertainties.

We would like to emphasize that our intention is not to advocate for a universal approach of ``rapid exploration'' in all asteroid missions. Instead, our objective is to illustrate the lack of necessity in minimizing uncertainties to an excessively low level for autonomous robotic spacecraft. We aim to demonstrate that autonomous robotic spacecraft possess the capability to effectively handle uncertainties, thus reducing the time spent solely on uncertainty reduction for navigation purposes. We fully recognize the significance of prolonged periods dedicated to sensor and hardware testing, calibration, detecting contingencies, extensive imaging from various phase angles, and other critical activities.

\section{Dynamical Environment}
\label{sec:dinamica}

Three reference frames are the most used for operations close to asteroids: inertial, fixed in orbit, and fixed in the body. The simplest is the inertial frame, centered in the asteroid and according to the ICRF (International Celestial Reference System). The frame fixed in the body is commonly defined with its $z$ axis pointing to the body's north pole, while the $x$ axis is in the direction of the smallest moment of inertia. Given the right ascension and declination of the pole of the asteroid in the inertial frame, respectively $RA$ and $DEC$, it is trivial to obtain a sequence of rotations that leads from the inertial system to the body-fixed:
\begin{align}
\label{eq:R_I_to_BF}
R_{I}^{BF} = R_z(\varpi) R_y(90\degree - DEC) R_z(90\degree+RA) .
\end{align}

The angle $\varpi$ represents a phase difference in the alignment of both systems' $x$ and $y$ axes. If the asteroid rotates uniformly about its largest principal axis of inertia (i.e., not considering wobble of the pole \cite{leonard2019osiris}), the angle $\varpi$ can be described as $\varpi=\varpi_0 + \dot{\varpi} \Delta t$, where the first term is usually defined with respect to the prime meridian at J2000~\cite{archinal2018report} (we make no ephemerides consideration on $\varpi$). The second is the spin angular velocity of the asteroid multiplied by the elapsed time. We define that $R_x(\theta)$, $R_y(\theta)$, and $R_z(\theta)$ are the three basic rotation matrices around the $x$, $y$, and $z$ axes, respectively, following the right-hand rule for the angle $\theta$ sign.  

The remaining frame of reference, orbit-fixed, is centered in the asteroid and defined with its $z$ axis in the direction of the angular momentum of the asteroid's heliocentric orbit, while the $x$ axis is in the direction of its heliocentric position. A transformation that takes from the orbit-fixed frame to the inertial one can be obtained using the osculating orbital elements of the asteroid's orbit, as:
\begin{align}
\label{eq:R_OF_to_I}
R_{OF}^{I} = R_z(-\Omega_a) R_x(-i_a) R_z(-\omega_a - \nu_a ) .
\end{align}
where $\Omega_a$, $i_a$, $\omega_a$, and $\nu_a$ represent, respectively, the osculating orbital elements: longitude of the ascending node, inclination, argument of periapsis, and true anomaly.


The equations of motion used in this study are in the orbit-fixed frame. If, as an approximation, the asteroid's orbit is assumed as an unperturbed two-body problem with the sun, and considering the orbital angular velocity of the asteroid, $\dot{\vec{\nu}}_a$, the equations of motion are~\cite{scheeres2016orbital,takahashi2021autonomous}:
\begin{subequations}
\label{eq:eqs_of_motion}
\begin{align}
\dot{\vec{r}} &=  \vec{v}, \\
\dot{\vec{v}} &= - \ddot{\nu}_a (\hat{z} \times \vec{r}) - \dot{\nu}_a ( 2 \hat{z} \times \vec{v} ) - \dot{\nu}_a [ \hat{z} \times ( \hat{z} \times \vec{r} )]+ \vec{a},
\end{align}
\end{subequations}
in which $\vec{r}$ and $\vec{v}$ stand for the position and velocity of the spacecraft in the orbit-fixed frame, and $\vec{a}$ represents the accelerations due to the forces acting on the spacecraft. The dot signs over the vectors are time derivatives. From the two-body problem, one easily finds that~\cite{takahashi2021autonomous}:
\begin{subequations}
\begin{align}
\dot{\nu}_a &= (1+e_a \cos \nu_a )^2 \sqrt{ \frac{\mu_S}{[a_a (1-e_a^2)]^3} }, \\
\ddot{\nu}_a &= - 2 e_a   \sqrt{ \frac{\mu_S}{[a_a (1-e_a^2)]^3} } \sin \nu_a (1+e_a \cos \nu_a ) \dot{\nu}_a ,
\end{align}
\end{subequations}
for $a_a$ representing the semi-major axis of the asteroid, $e_a$ its eccentricity, and $\mu_S$ the gravitational parameter of the sun.

Due to the asteroid's low gravity, many different forces reasonably affect the spacecraft operating in its vicinity. The importance of considering or not the action of a particular force varies from case to case, and it should be thought of according to the objective of the analysis, distance to the asteroid, mass, and other properties. In general, as exemplified in Antreasian et al. \cite{antreasian2016osiris} for the OSIRIS-REx mission, the main force when the spacecraft identifies the asteroid in its optical cameras, hundreds of kilometers away, is the solar radiation pressure (SRP), third-body gravitational effects from the sun, and the central term of the asteroid's gravity field. At a few asteroid diameters of distance to its center of mass, the spacecraft starts to experience gravitational effects due to the non-uniformity of its volume and density, which end up surpassing or equalling the SRP and third body effects. The asteroid's albedo and its infrared radiation pressure also have a considerable effect.

We consider all of these main forces, except albedo and infrared radiation pressure (due to their complexity, dependent on the asteroid's shape and composition), with the addition of accelerations from the spacecraft thrusters performed by the control system. Therefore, the acceleration $\vec{a}$ on the spacecraft in the chosen frame of reference is:
\begin{equation}
\label{eq:accelerations}
\vec{a} = \vec{a}_{g} + \vec{a}_{SRP} + \vec{a}_{3B} + \vec{u},
\end{equation}
The subscripts $g$, $SRP$, and $3B$ represent the gravitational field of the asteroid, the SRP, and the third-body effects due to the sun. The vector $\vec{u}$ is the control command.

For solar radiation pressure, a cannonball model is assumed, given by:
\begin{equation}
\label{eq:SRP}
\vec{a}_{SRP} = \frac{P_0 (1+\rho) U^2 A_{SC}}{M_{SC}} \frac{\vec{r}-\vec{d}}{ \vert \vert \vec{r}-\vec{d} \vert \vert ^3 },
\end{equation}
where $\vec{r}$ is the position of the spacecraft in the considered frame of reference, $\vec{d}$ is the position of the sun relative to the asteroid, $\rho$ is the reflectivity of the surface, $P_0= 4.56\times 10^{-6}$ N/m$^2$, $U$ is the distance of 1 AU in meters, $A_{SC}$ the considered surface area and $M_{SC}$ the mass of the spacecraft. In the simulations the considered values are\cite{takahashi2021autonomous}: $A_{sc}=16$ m$^2$, $M_{sc}=1000$ kg, and $\rho=0.4$.

In the case of the sun's gravitational effects, we have:
\begin{equation}
\vec{a}_{3B} = -\mu_S \left( \frac{\vec{d}}{\vert \vert \vec{d} \vert \vert ^3} + \frac{\vec{r }-\vec{d}}{\vert \vert \vec{r}-\vec{d} \vert \vert ^3 } \right).
\end{equation}
 
The gravitational field of the asteroid will be simulated using a spherical harmonic expansion, which is obtained considering the gravitational potential:
\begin{equation}
\label{eq:spher_harm}
    U_{sh} = \frac{\mu}{r} \sum^{N}_{n=0} \sum^n_{m=0} \left( \frac{r_0}{r} \right)^n P_{nm}(\sin \varphi) \left[ C_{ nm} \cos{(m\varrho)} + S_{nm} \sin{(m\varrho)} \right],
\end{equation}
where $\varphi$ and $\varrho$ are, respectively, the latitude and longitude of the spacecraft in the body-fixed frame and $P_{nm}$ are the associated Legendre polynomials. Recursive formulae to calculate the potential are used, which also obtain the acceleration $\vec{a}_{g}^{BF}$ \cite{montenbruck2002satellite}. The spherical harmonics expansion is made up to the fifth order and degree. 

For this study, we have selected two asteroids: Bennu and Eros. The specific parameters of these asteroids used in our simulations are provided in Table \ref{tab:Itokawa}. When the spacecraft enters Brillouin's sphere at any given moment in the simulation, a polyhedron model is employed to describe the gravity field while inside the sphere. The acceleration $\vec{a}_{g}^{BF}$ is determined in the body-fixed frame using the polyhedron gravity model \cite{werner1996exterior}, assuming a constant density of $\delta=1177.05$ kg/m$^3$ for Bennu and $\delta=2681.77$ kg/m$^3$ for Eros. 

For the spherical harmonic expansion, the coefficients $C_{nm}$ and $S_{nm}$ in Equation \ref{eq:spher_harm} are calculated based on the polyhedron representation of the asteroid's shape \cite{werner1997spherical}, ensuring consistency with the assumed shape characteristics. The shape data for the asteroids is obtained from NASA's PDS Small Bodies Node website\footnote{\url{https://sbn.psi.edu/pds/shape-models/}} and processed using Autodesk Meshmixer to reduce it to 4,000 facets. It should be noted that the body-fixed frame is positioned at the asteroid's center of mass and aligned with its principal axes of inertia based on the assumed mass distribution \cite{dobrovolskis1996inertia}.

We have purposely selected those specific asteroids to underscore the fact that our proposed guidance, navigation, and control ($GN\&C$) approach is not reliant on the size or shape of the asteroid. The mission profile can be customized based on the specific objectives of the mission and the available information about the asteroid's environment and properties. Once the initial assessment of the environment is conducted, the spacecraft can consider various profiles based on the overall characteristics of the asteroid as observed from a distance. This preliminary analysis aids in determining whether solar radiation pressure or the asteroid's elongated shape is the predominant factor influencing the mission. For scenarios where solar radiation pressure dominates, a sun-terminator orbit would be suitable, while for elongated asteroids, a retrograde equatorial orbit in the asteroid's inertial frame would be generally preferable~\cite{scheeres2012orbit,scheeres2014close,scheeres2016orbital}. However, it is important to note that selecting the most suitable approach may vary depending on the specific mission objectives and the characteristics of the individual asteroid.

\begin{table}[!h]
	\fontsize{10}{10}\selectfont
    \caption{Parameters of the Asteroids.}
   \label{tab:Itokawa}
        \centering 
   \begin{tabular}{c  c  c  c  c  c  c  c  c } 
   \hline
   	  Asteroid &
      $\mu$ [m$^3$/s$^2$]      & 
      $DEC$ [$\degree$]        & 
      $RA$  [$\degree$]       & 
      $a_a$ [AU]  & 
      $e_a$   & 
      $i_a$ [$\degree$]   &  
      $\omega_a$ [$\degree$]   & 
      $\Omega_a$ [$\degree$]  \\
      \hline
      \begin{tabular}{@{}c@{}} \\  Bennu \end{tabular} & \begin{tabular}{@{}c@{}} \\  4.88844 \end{tabular} & \begin{tabular}{@{}c@{}} \\ -60.36 \end{tabular} & \begin{tabular}{@{}c@{}} \\ 85.46 \end{tabular} & \begin{tabular}{@{}c@{}} \\ 1.1264 \end{tabular} & \begin{tabular}{@{}c@{}} \\ 0.2037 \end{tabular} & \begin{tabular}{@{}c@{}} \\ 6.0349 \end{tabular} & \begin{tabular}{@{}c@{}} \\ 66.2231 \end{tabular} & \begin{tabular}{@{}c@{}} \\ 2.0608 \end{tabular} \\
      \begin{tabular}{@{}c@{}} \\  Eros \end{tabular} & \begin{tabular}{@{}c@{}} \\   $4.46023\times 10^{5}$ \end{tabular} & \begin{tabular}{@{}c@{}} \\ 17.22 \end{tabular} & \begin{tabular}{@{}c@{}} \\ 11.35 \end{tabular} & \begin{tabular}{@{}c@{}} \\ 1.4583 \end{tabular} & \begin{tabular}{@{}c@{}} \\ 0.2228 \end{tabular} & \begin{tabular}{@{}c@{}} \\ 10.8292 \end{tabular} & \begin{tabular}{@{}c@{}} \\ 178.6653 \end{tabular} & \begin{tabular}{@{}c@{}} \\ 304.4010 \end{tabular} \\
      \hline
   \end{tabular}
\end{table}

\section{Navigation}

\begin{figure}
	\centering\includegraphics[width=1\textwidth]{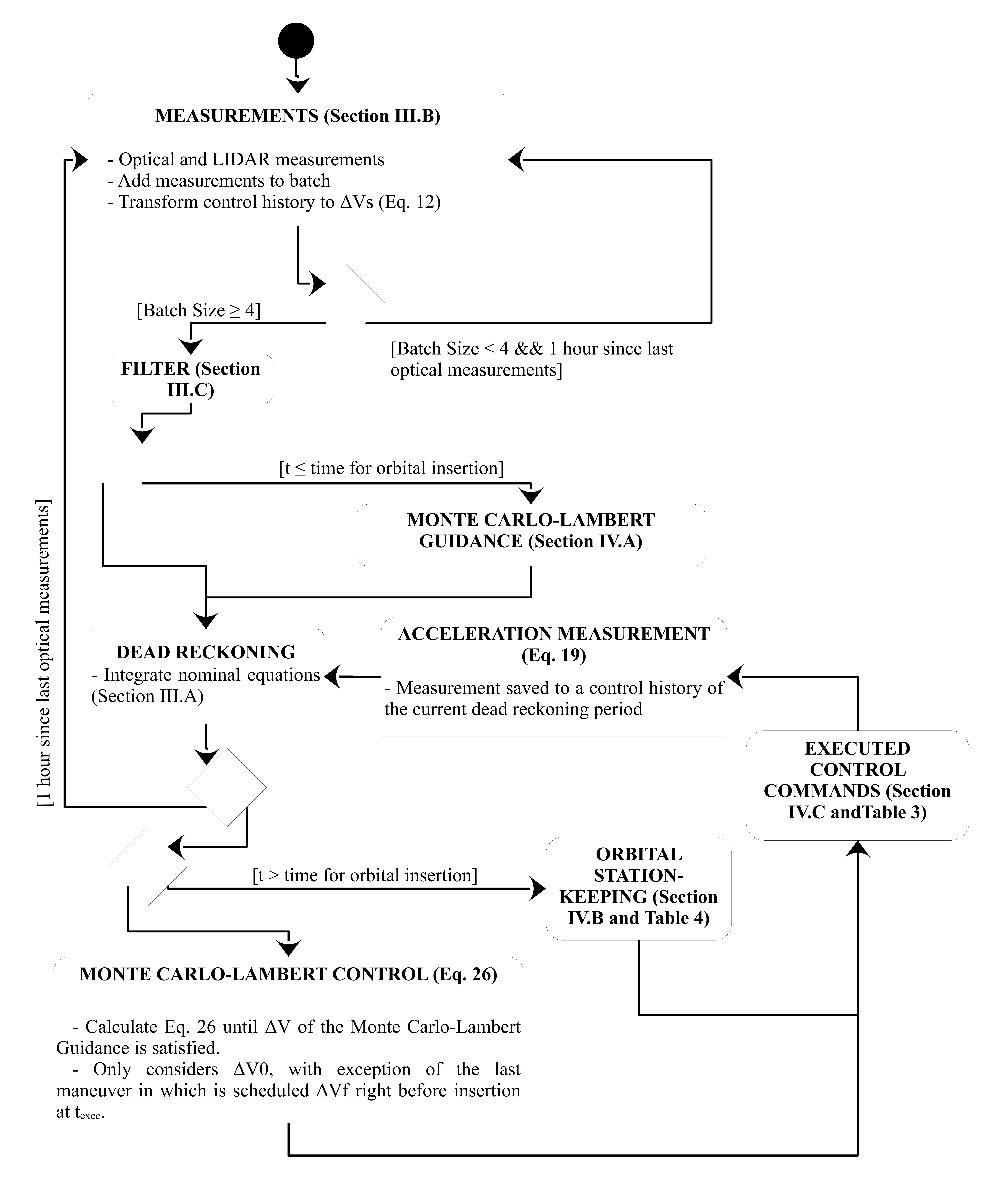}
	\caption{Finite-State Machine of the proposed GN$\&$C flight software within the simulation.}
	\label{fig:state_machine}
\end{figure}

In this section, we describe the navigational solutions applied for this proposed GN\&C (guidance, navigation, and control) architecture. A finite-state machine is depicted in Fig. \ref{fig:state_machine} to aid the reader in understanding the relation between different aspects of the GN\&C codes and the simulation. First, assume a non-linear system of differential equations, with an output $\vec{Y}$, written as:
\begin{subequations}
\label{eq:filter_nonlinear}
\begin{align}
\dot{\vec{X}} &= \vec{F}(\vec{X},t), \\
\vec{Y} &= \vec{G}(\vec{X},t) + \vec{\varepsilon},
\end{align}
\end{subequations}
$\vec{X} \in \mathbb{R}^n$, $\vec{Y} \in \mathbb{R}^p$, and $\vec{\varepsilon} \in \mathbb{R}^p$ representing uncertainties introduced to $\vec{Y}$. The problem is then of estimating the state vector $\vec{X}$ for a set of measurements $\vec{Y}$.

\subsection{Onboard Nominal Dynamics and Environment}
\label{sec:nominal}

It is considered as the state vector in Equation \ref{eq:filter_nonlinear}:

\begin{equation}
\vec{X} = \begin{bmatrix}
\vec{r} & \vec{v} & \mu & C_{SRP}
\end{bmatrix}^\text{T},
\end{equation}
in which $C_{SRP}$ includes all the constant terms in Equation \ref{eq:SRP}. The choice in not including any information about the asteroid's shape (e.g., its mean radius as used in Reference \cite{takahashi2021autonomous}), spin state, $\Delta \vec{V}$s applied by the control system, or higher-order terms of the gravity field is deliberate. We assume that in the preliminary phase of the approach, the spacecraft (autonomously or not) has obtained an accurate estimate of the asteroid's attitude and spin state. This is a reasonable assumption if noted the achievements in the OSIRIS-REx campaign when the spacecraft was 162 km away from Bennu, with errors in the order of $0.2 \degree$ and lower for $RA$, $DEC$, and $\dot{\varpi}$ \cite{leonard2019osiris}. 

In the nominal dynamics~\footnote{The term ``nominal dynamics'' refers to the model of the dynamical environment embedded in the spacecraft's computer. Robotic spacecraft will likely not possess highly complex embedded models describing any small force acting on the spacecraft, due to adding unnecessary complexity to the flight software. In some cases, the environment may also be unknown.} is only considered the central term of the gravity field, i.e. $\vec{a}_g=-\frac{\mu}{r^3} \vec{r}$ in Equation \ref{eq:accelerations}. The choice of neglecting higher-order terms in the nominal gravity field, and thus not estimating them in $\vec{X}$, is to stress the GN\&C proposal in the case of unmodeled dynamics. The terms $\mathcal{J}_2$ and $C_{22}$ of the asteroids' gravity field will have a huge impact on the dynamics of the spacecraft operating in their proximity, with almost two orders of magnitude larger than the SRP. Therefore, we expect that neglecting the higher-order terms in the nominal scenario and estimates will provide a conservative assessment of other neglected effects such as infrared radiation pressure, albedo, and attitude-dependent SRP. {This is especially true for Eros, chosen for being both highly elongated and massive, to rigorously stress and assess the proposition robustness.}

The OSIRIS-REx, when operating close to Bennu, relied on a shape model obtained through stereophotoclinometry (SPC) with a 3D error of 1 m \cite{al2021validation}. This accuracy is equivalent to about 0.4$\%$ of Bennu's mean radius. Although reliable and accurate, this shape reconstruction approach is time-consuming and computationally intensive. Less demanding approaches are still an area of active research, primarily to facilitate onboard shape reconstruction \cite{bercovici2019robust,baker2020limb}. Due to these limitations, and to not employ an arguably quite optimistic scenario for the shape model as was done in previous works \cite{takahashi2021autonomous}, we assume that instead of continuously estimating the asteroid's shape, the spacecraft has a rough onboard shape with an assumed uncertainty level.

Finally, the reason for not including $\Delta \vec{V}$s in the state vector is the high activity of the control system during the fast approach to the asteroid with significant uncertainties. If the $\Delta \vec{V}$s were included in $\vec{X}$, it would be unclear whether the spacecraft could handle a large $\vec{X}$ between its optical and LiDAR measurements. Therefore, a conservative approach is taken where the measured control commands are passed as parameters to the nominal equations. This approach takes two forms.

In the first approach, the spacecraft integrates the nominal equations using the most recent estimation of $\vec{X}$ during the periods between optical measurements. This integration takes place in real-time over a time interval equivalent to the update time of the control commands, denoted as $\Delta t_{u}$. As a result, the control commands are considered constant values in the equations of motion. The control system relies exclusively on the IMU data for measurements during this phase, which is commonly referred to as dead reckoning. This can be observed in the state labeled ``dead reckoning'' in Fig. \ref{fig:state_machine}, where the spacecraft utilizes current acceleration measurement from the IMU to propagate its trajectory in real time.

The second way the control commands are processed is within the filter, where they are treated as if they were set of $\Delta \vec{V}$s in a time-dependent function:

\begin{equation}
\label{eq:dVs}
 \vec{u}(t) = \sum_{j=1}^{N_{\Delta V}} \Delta \vec{V}_j \delta(t-t_j),
\end{equation}
where $\delta(t-t_j)$ is the Dirac delta function, and $\Delta \vec{V}_j$ is a $\Delta \vec{V}$ applied at time $t_j$. In practice, when integrating the equations of motion in the estimation, the $\Delta \vec{V}_j$ is simply added to the velocity in the state at times $t_j$ (more details in Section \ref{sec:estimation}).

The way the control commands $\vec{u}$ are reduced to Equation \ref{eq:dVs} is very simple. The control command history is stored between the optical measurement times to be later fed into a routine. That routine, before the filter initialization, analyzes the control history and reduces different commands within time intervals $\Delta t_{\Delta V}=60$ s to each $\Delta \vec{V}_j$:
\begin{equation}
\Delta \vec{V}_j = \sum_{t_j \leq t < t_j + \Delta t_{\Delta V}} \Delta t_{\Delta V} \vec{u}(t),
\end{equation}
as represented in the state ``measurements'' in Fig. \ref{fig:state_machine}.

{

We note that this routine to reduce the control commands to $\Delta \vec{V}_j$s will need to analyze only the history in the last dead-reckoning period. Once a reduction is made, the $\Delta \vec{V}_j$s will be probably stored in the read-only memory (ROM), most likely an EEPROM (Electrically Erasable Programmable ROM). The reduced control commands will also likely be kept in RAM (random-access memory) for low-latency access while the filter needs them in its later runs. We see no reason to believe that typical access times found in space-graded RAM chips will have a substantial effect on the autonomous GN\&C system operation. Typical control history in a dead-reckoning period will only be a few bytes and it will likely not be a source of concern either.
}

\subsection{Measurements}
\label{sec:measu}

We assume that the spacecraft is equipped with a LiDAR (Light Detection and Ranging), two optical navigation cameras and a set of accelerometers, for navigation with respect to the asteroid. A summary of the values used in the simulation is presented in Table \ref{tab:variable_values}. We consider that no radiometric data is available for the spacecraft, as one of the significant advantages of an autonomous operation is alleviating the ground burden on operating the spacecraft and in removing trajectory constraints that would need to be placed for maintaining a regular telemetry. The LiDAR measurements are modelled following the relation for their accuracy error as:

\begin{equation}
\sigma_{LiDAR} = \begin{cases}
    5.5\text{ m},& \text{if } r\geq 6\text{ km}\\
    0.1\text{ m},& \text{if } \text{otherwise}
\end{cases},
\end{equation}

We note that this is a conservative approach. Advanced Scientific Concepts, the company responsible for the OSIRIS-REx 3D flash LiDAR used in the GN\&C, reports a range error of 5-10 cm for the model ``GSFL-16KS Space''\footnote{\url{https://asc3d.com/gsfl_16Ks/}}, in a range below 6 km. Mizuno et al. \cite{mizuno2017development} report for Hayabusa 2 an operational range from 30 m to 25 km, with 1 m to 5.5 m errors at each respective range bound. Although both LiDARs are very different (e.g., Hayabusa 2 LiDAR is not a 3D flash LiDAR), we think it is reasonable to mix both to avoid taking an over-conservative approach. As we will show soon, the range measurements are assumed to be made only by the LiDAR, which is unrealistic. In practice, the optical navigation would complement the range data settling the significance of the LiDAR errors if there is already an onboard shape of the asteroid. In fact, in some missions, as is the case of OSIRIS-REx, the LiDAR measurements are not even considered in the orbit determination, being more critical for fault detection~\cite{williams2018osiris}.

\begin{table}[!h]
\centering
\caption{Summary of assumed values in the navigation.}
\label{tab:variable_values}
\begin{tabular}{ll}
\hline
\textbf{Variable} & \textbf{Assumed value} \\
\hline
LiDAR uncertainty ($\sigma_{LiDAR}$) &
$\begin{cases}
5.5  \text{ m}, & \text{if } r \geq 6 , \text{km} \\
0.1  \text{ m}, & \text{otherwise}
\end{cases}$ \\
Wide $FOV$ camera & 69.71$\degree$ \\
Narrow $FOV$ camera & 6.27$\degree$  \\
Cameras' pixels ($N_p$) & 1024$\times$1024 pixels \\
Shape model uncertainty ($\sigma_R$) & 1\% \\
IMU uncertainty ($\sigma_{IMU}$) & $3.4 \times 10^{-4}$ m/s$^2$ \\
Thruster's transfer function uncertainty ($\sigma_{TF}$) & 1\% of the executed control \\
Cadence of optical and LiDAR measurements & 1 hour \\
\hline
\end{tabular}
\end{table}

Hayabusa 2 spacecraft has three optical navigation cameras, ONC-T, ONC-W1, and ONC-W2~\cite{takei2020hayabusa2}. For the optical navigation of our analysis scenario, we consider the ONC-T and ONC-W1. The ONC-T is a telescopic camera with a FOV of 6.27$\degree$ and pixel size of 1024$\times$1024, while the ONC-W1 has a wide FOV of 69.71$\degree$ and 1024$\times$1024 pixels.

The optical and LiDAR measurements are combined to form the measurement vector $\vec{Y}$ that are fed into the navigation filter to estimate $\vec{X}$:
\begin{equation}
\vec{Y} = \begin{bmatrix}
r & \hat{r}
\end{bmatrix}^\text{T}
\end{equation}

For the range measurement $r$, its uncertainty is assumed as:

\begin{equation}
\sigma_r^2 \sim  \sigma_{LiDAR}^2 + \sigma_R^2 R^2,
\end{equation}
in which $R$ is an assumed size of the body and $\sigma_R$ is due to shape uncertainties. As already discussed, note that this is a conservative approach as silhouette-based techniques can be employed to complement and further constrain the range data if there is already a shape model of the asteroid with known uncertainty, decreasing the significance of $\sigma_{LiDAR}$ \cite{liounis2018limb}.

We assume a procedure similar to Reference \cite{scheeres2019autonomous} for obtaining $\hat{r}$. Assuming two angles $\theta_1$ and $\theta_2$ coming from comparing the scene obtained from the optical cameras with the star background, we define:
\begin{equation}
\hat{r} = \begin{bmatrix}
\cos \theta_1 \cos \theta_2 \\
\sin \theta_1 \cos \theta_2 \\
\sin \theta_2
\end{bmatrix}.
\end{equation}
With this definition, we can model uncertainties in the optical measurements according to~\cite{scheeres2019autonomous}:
\begin{equation}
\sigma_\theta^2  \sim  IFOV^2 + \sigma_\Omega^2,
\end{equation}
in which $IFOV = FOV / N_p$ is the instantaneous field of view of the camera, with $N_p=1024$ being the number of pixels. The term $\sigma_\Omega$ represents the angular uncertainty of the asteroid, represented as~\cite{scheeres2019autonomous}:
\begin{equation}
 \sigma_\Omega^2 \sim  \frac{\sigma_R^2 R^2}{r^2},
\end{equation}
where $\sigma_R$ is the 3D error in percentage in the shape model with respect to a reference size $R$, in which we take the mean radius of the polyhedron model that we use as the real asteroid ($R=246.8$ m for Bennu, and $R=8795.0$ m for Eros). As already argued, $\sigma_R$ is within 0.4$\%$ in the SPC shape reconstruction of Bennu. Scheeres \& McMahon \cite{scheeres2019autonomous} argue that current research is on the way to constrain $\sigma_R$ by 1$\%$. In this study, we then mostly apply $\sigma_R=1\%$.

Takahashi $\&$ Scheeres \cite{takahashi2021autonomous} assume the use of accelerometers with a noise spectral density of $1\times 10^{-5}$ m/s$^2$/$\sqrt{\text{Hz}}$ for measuring the applied $\Delta \vec{V}$ maneuvers, an order of magnitude more precise than off-the-shelf inertial measurement units (IMU). They argue that state-of-the-art accelerometers have a noise spectral density below $1\times 10^{-8}$ m/s$^2$/$\sqrt{\text{Hz}}$, which would make their assumption proper. However, we note that these ultra-precise state-of-the-art accelerometers have an operational bandwidth that is not adequate for a spacecraft's attitude and orbit control subsystem (AOCS). They are designed to operate at a maximum frequency of 0.1 Hz, as can be checked in the accelerometers' table presented in Takahashi $\&$ Scheeres and references therein \cite{takahashi2021autonomous}. 

Here we take what we consider a more suitable approach by assuming the use of an off-the-shelf IMU, for which we choose the LN-200S \footnote{\url{https://www.northropgrumman.com/what-we-do/ln-200s-inertial-measurement-unit/}}, the same used in the Hayabusa 2 mission, that has a noise spectral density of $3.4\times 10^{-4}$ m/s$^2$/$\sqrt{\text{Hz}}$. We also assume that any small-thrust engine equipped on the spacecraft (hence producing a thrust level in the same order or below the chosen IMU's noise) has passed through a meticulous campaign to obtain their transfer function so that other measurements, such as pressure in the chamber, current, and others (depending on the engine), can be related to the thrust level. 

In that way, we assume a conservative 1$\sigma$ uncertainty of $1\%$ in this indirectly measured thrust level. That uncertainty level is in the same order as the throttle curves obtained for the SPT-140 thrusters used in the Psyche mission~\cite{snyder2019electric}, and much greater than the 0.15$\%$ noise of the micro-thrusters used in the LISA Pathfinder~\cite{tajmar2004indium}. Therefore, both sources of thrust measurement, direct (IMU) and indirect, can be fused to obtain the imparted thrust level. The measured thrust is then obtained from the sensor fusion:

\begin{equation}
    u_{\text{meas}} = \frac{\sigma_{TF}^{-2}u_{TF}+\sigma_{IMU}^{-2}u_{IMU}}{\sigma_{TF}^{-2}+\sigma_{IMU}^{-2}},
\end{equation}
which is simply an inverse-variance weighting, where $\sigma_{IMU} = 3.4\times 10^{-4}$ m/s$^2$ considering a 1 Hz measurement frequency, $\sigma_{TF} = 1\%$ of the actually executed control, $u_{TF}$ is the thrust level obtained through the thrusters' transfer function, and $u_{IMU}$ is the thrust level measured by the IMU. From the inverse-variance weighting method, we know that the $1\sigma$ uncertainty in the measured control is $\sqrt{1/(\sigma_{TF}^{-2}+\sigma_{IMU}^{-2})}$.

\subsection{Estimation}
\label{sec:estimation}

For the estimation process, we employ the same filtering approach adopted in the JPL's (Jet Propulsion Lab) AutoNav \cite{bhaskaran1998orbit,riedel2000autonomous,riedel2006autonav,bhaskaran2012autonomous}. It is a batch-sequential least-squares filtering approach, in which the measurements are processed as a batch that sequentially advances in time as new measurements are made. The choice of that approach is because it presents better stability than a sequential one for the case of sparse measurements, as it uses a consistent reference trajectory~\cite{takahashi2021autonomous}. In addition, the possibility of data editing to eliminate outliers and the ability to identify systematic anomalies give an additional degree of robustness to the filter \cite{riedel2006autonav,bhaskaran2012autonomous}, even though we make no consideration in that regard. Its limitation is a larger required processing time if compared to the sequential process. However, there is plenty of time for processing a reasonable batch of data in an asteroid mission due to the sparse measurements. An indication of that fact is the measurement intervals in the order of hours in the asteroid missions. For instance, the OSIRIS-Rex mission makes a new measurement every 2 hours \cite{williams2018osiris}. Here, we define that a new measurement is made each hour.

Let the error between the real and the reference trajectory be defined, considering a discretization for the times $t_i$, $i=1,2,...,l$, as $\bar{\vec{x}}_i=\vec{X}_i-\vec{X}_i^*$, in which $l$ is the size of the batch. Likewise: $\bar{\vec{y}}_i=\vec{Y}_i-\vec{Y}^*_i$. Expanding the discretized Equations \ref{eq:filter_nonlinear} to $t_i$ in Taylor series to obtain a linear approximation, we find~\cite{schutz2004statistical}:
\begin{subequations}
\label{eq:filter_linear}
\begin{align}
\dot{\bar{\vec{x}}}_i &= A_i(t_i)\bar{\vec{x}}_i, \\
\bar{\vec{y}}_i &= C_i(t_i)\bar{\vec{x}}_i + \vec{\varepsilon}_i,
\end{align}
\end{subequations}
in which:
\begin{subequations}
\begin{align}
A_i(t_i) & = \left. \frac{\partial \vec{F}}{\partial \vec{X}} \right\vert_{\vec{X}_i=\vec{X}^*_i,t=t_i}, \\
C_i(t_i) & = \left. \frac{\partial \vec{G}}{\partial \vec{X}} \right\vert_{\vec{X}_i=\vec{X}^*_i,t=t_i}.
\end{align}
\end{subequations}

We want to estimate for an epoch $t_k$, which we choose as the time of the last measurement ($k=l$), to obtain $\bar{\vec{x}}_k$. So it is now possible to use the solution of a linear system:
\begin{subequations}
\label{eq:filter_int}
\begin{align}
\bar{\vec{x}}_i &=\Phi(t_i,t_k )\bar{\vec{x}}_k, \\
\dot{\Phi}(t,t_k ) &= A(t) \Phi(t,t_k ),
\end{align}
\end{subequations}
where $\Phi(t,t_k )$ is the state transition matrix, for $\Phi(t_0,t_0) = I$. Now, we can transform Equation \ref{eq:filter_linear} into:
\begin{subequations}
\begin{align}
\dot{\bar{\vec{x}}}_i &= A_i(t_i)\Phi(t_i,t_k)\bar{\vec{x}}_k, \\
\bar{\vec{y}}_i &= C_i(t_i)\Phi(t_i,t_k)\bar{\vec{x}}_k + \vec{\varepsilon}_i.
\end{align}
\end{subequations}

Defining $\vec{\varepsilon} = \begin{bmatrix}\vec{\varepsilon}_1 & \vec{\varepsilon}_2 & ... & \vec{\varepsilon}_k\end{bmatrix}^\text{ T}$ and assuming a normal distribution with zero mean, $E[\vec{\varepsilon}]=0$, and known covariance, $E[\vec{\varepsilon}\vec{\varepsilon}^\text{T }]=R$, it is possible to demonstrate that the estimate with minimum covariance for $\bar{\vec{x}}_k$ is~\cite{schutz2004statistical}:
\begin{equation}
\label{eq:xk_est}
\hat{\bar{\vec{x}}}_k = (C^\text{T}R^{-1}C)^{-1}C^\text{T}R^{-1}\bar{\vec{y}},
\end{equation}
where $C = \begin{bmatrix}C_1(t_1)\Phi(t_1,t_k) & C_2(t_2)\Phi(t_2,t_k) & ... & C_k(t_k)\end{bmatrix}^\text{T}$ and $\bar{\vec{y}}=\begin{bmatrix}\bar{\vec{y}}_1 & \bar{\vec{y}}_2 & ... & \bar{\vec{y}}_k\end{bmatrix}^\text{T}$. {The states' covariance matrix for the epoch is $P_k=(C_k^\text{T}R_k^{-1}C_k)^{-1}$. 

Equation \ref{eq:xk_est} is iteratively solved following the algorithm in Reference \cite[pg. 196]{schutz2004statistical} up to convergence of the root-mean-square (RMS) of the observation residuals to obtain the epoch's estimate $\hat{\vec{X}}_k$. We note that the integration of Equations \ref{eq:filter_int} is stopped in the times $t_j$ to properly take into account the $\Delta \vec{V}_j$s, as shown in Equation \ref{eq:dVs}, in the components of $\vec{X}$ correspondent to $\vec{v}$.

The epoch's covariance matrix can be propagated in time to be used in the next estimation cycle as:
\begin{equation}
    \hat{P}_{k+} = W \Phi(t_{k},t_{k+}) P_{k} \Phi^\text{T}(t_k,t_{k+}).
\end{equation}
When applying the batch estimator to non-linear systems with unmodeled dynamics, the primary utility of the estimate $\hat{P}_{k+}$ lies in enhancing the conditioning of the covariance matrix, rather than offering an accurate representation of uncertainties on the new $\bar{\vec{x}}_k$~\cite[pg. 198]{schutz2004statistical}. Therefore, the parameter $W$ is often used in these cases as a weighting matrix, in a process commonly known as covariance inflation. The same procedure is used in JPL's AutoNav~\cite{riedel2000autonomous}. 

In this study, we will take only the diagonal of $\Phi(t_{k},t_{k+}) P_{k} \Phi^\text{T}(t_k,t_{k+})$ and make the weighting parameter as a scalar $W=4$. The engineers of JPL have reported the use of $W=2$ in an application similar to ours~\cite{nesnas2021autonomous}. However, we impose a lot of what can be considered over-conservative approaches \footnote{For instance, incorporating the second-order term of the gravity field of a highly elongated and massive body like Eros into the ``unmodeled dynamics'' serves as a test to assess the robustness of the proposition, check Section \ref{sec:nominal}.}, so a larger value for $W$ is adequate. Larger covariance inflation means a more conservative approach to the estimation, which generally~\footnote{We employ the term ``generally'' because, in the event of excessively large covariance inflation, estimation errors are likely to escalate, potentially destabilizing the system.} brings more robustness to the operation, at the cost of decreasing the estimation precision. In the first round of estimation, we assume no uncertainty is available. This is equivalent to making the information matrix~\footnote{The information matrix is the inverse of the covariance matrix.} a zero matrix, which is the most conservative approach from an estimation perspective.
}

\subsection{Considerations on Onboard Computation}
 
Advancements in embedded technology have significantly expanded the capabilities of spacecraft for performing complex flight algorithms. This is exemplified by the continuous improvement of processors and their ability to handle demanding computational tasks. JPL's AutoNav system, successfully deployed in missions like Deep Impact and Deep Space 1, serves as a solid foundation to support the feasibility of complex computations onboard a spacecraft~\cite{bhaskaran1998orbit,bhaskaran2012autonomous,bhaskaran2020autonomous}. For instance, the Deep Impact mission employed the BAE Systems Electronics RAD750 processor~\footnote{\url{https://web.archive.org/web/20070711152307/http://www.ballaerospace.com/page.jsp?page=96}}, commonly used in JPL's missions, which can operate at 200 MHz. Despite AutoNav consuming a substantial portion of CPU resources, approximately half of the CPU cycles~\cite{riedel2006autonav}, it exemplifies the capability of flight-proven technology to manage intricate computing tasks. 

The successor to the Systems on a Chip (SoCs) employing the RAD750 is the RAD510 SoC, featuring the RAD5500 processor core, which represents a significant advancement in embedded processing from RAD750. The RAD5500 processor core operates within a frequency range of 66 MHz to 462 MHz, enabling it to cater to both low-power and high-performance applications~\footnote{\url{https://www.baesystems.com/en/our-company/our-businesses/electronic-systems/product-sites/space-products-and-processing/radiation-hardened-electronics-produ}}.

Further underscoring the notion that processing power is no longer a limiting factor in space exploration, we turn to the LEON4 processor developed by Aeroflex Gaisler for the European Space Agency (ESA). Operating at clock frequencies of up to 1500 MHz on 32 nm ASIC (Application-Specific Integrated Circuit) technologies, the LEON4 processor delivers substantial computational power, surpassing its predecessors in terms of performance~\footnote{\url{https://www.esa.int/Enabling_Support/Space_Engineering_Technology/Onboard_Computers_and_Data_Handling/Microprocessors}}. Gaisler's GR740 SoC integrates four LEON4 cores operating at 250 MHz~\footnote{\url{https://www.gaisler.com/index.php/products/components/gr740}}, while the ongoing development of the GR765 SoC will feature eight cores of the LEON5 processor, capable of operating at 1 GHz~\footnote{\url{https://gaisler.com/index.php/products/components/gr765}}. 

These technological advancements highlight the significant progress made in embedded processing for spacecraft. The enhanced computational power of processors like RAD5500, LEON4, and the upcoming LEON5 offers new possibilities for executing complex computations efficiently and accurately onboard spacecraft, such as the filter shown in Section \ref{sec:estimation} and the Monte Carlo-Lambert guidance we will discuss in Section \ref{sec:lamb_guid}.

{The principal challenge in terms of onboard computation is expected to arise from the optical navigation algorithms, particularly those related to image processing. In the Deep Space 1 mission, utilizing the RAD6000 operating at 33 MHz, it was reported that the AutoNav system employed a streamlined version of navigation processing during its flybys near the small bodies, achieving completion within a brief timeframe of 10 to 15 seconds~\cite{riedel2000autonomous}. The integration of silhouette-based navigation into AutoNav could potentially intensify the demand for processing resources and necessitate the adoption of more intricate image-processing algorithms. 

Nonetheless, as previously noted, contemporary radiation-hardened processors surpass the clock frequency of RAD6000 by more than tenfold. Additionally, there exists the possibility of implementing image processing algorithms in field-programmable gate arrays (FPGAs), offering the potential for significant performance enhancements. Recent advancements in space-grade FPGAs, exemplified by AMD Xilinx's Kintex UltraScale~\footnote{https://www.xilinx.com/products/silicon-devices/fpga/rt-kintex-ultrascale.html}, further contribute to the feasibility of such implementations. Admittedly, the incorporation of more complex algorithms entails higher development costs, yet these investments are anticipated to yield long-term benefits by reducing reliance on ground-based spacecraft operations. The hardware implementation, for hardware-in-the-loop simulations, based on an architecture similar to that presented in this work, should be pursued in a more advanced phase of this research, with a thorough consideration of the optical navigation algorithms.}

\section{Guidance and Control}
\label{sec:G&C}

As already discussed, once an estimate $\hat{\vec{X}}_k$ for the epoch is obtained, it is used to propagate the nominal equations of motion before the subsequent measurement and estimate update (i.e., the spacecraft is in dead reckoning). The nominal equations of motion are propagated in real-time in intervals equal to the control cadence $\Delta t_u$, which we choose to be 1 second. We only use two different control approaches. The first is a Monte Carlo-Lambert guidance responsible for driving the spacecraft to orbital insertion from far away from the asteroid. The second one is an orbit-keeping control responsible for maintaining and transiting between Keplerian orbits about the asteroid. Both of them can be considered robust approaches because they explicitly deal with uncertainties, the first by taking a stochastic approach and the second by using sliding-mode control~\cite{slotine1991applied}. The finite-state machine already shown in Fig. \ref{fig:state_machine} also represents the relation of the guidance and control algorithms with other parts of the simulation.

Consider Equations \ref{eq:eqs_of_motion}. Let us rewrite it as:
\begin{subequations}
\begin{align}
\dot{\vec{r}} &= \vec{v}, \\
\dot{\vec{v}} &= \vec{f} + \vec{u} + \vec{d},
\end{align}
\end{subequations}
in which $\vec{f}= \ddot{\nu}_a (\hat{z} \times \vec{r}) - \dot{\nu}_a ( 2 \hat{z} \times \vec{v} ) - \dot{\nu}_a [ \hat{z} \times ( \hat{z} \times \vec{r} )]+ \vec{a}_{3B} + \vec{a}_{g} + \vec{a}_{SRP} + \vec{a}_{3B}$ and $\vec{d} \in \mathbb{R}^3$ is disturbances acting in the system and known to be bounded by $\vert d_i \vert <D_i$, for $i=1,2,3$ representing each of its components.

\subsection{Monte Carlo-Lambert Guidance}
\label{sec:lamb_guid}

A common feature in an asteroid mission is the high uncertainty level of the spacecraft's belief on the exact asteroid properties, environment, and relative state before being at a few body's radii of distance. For that reason, the Hayabusa 2 mission employs stochastic guidance for the approach phase, using the software JATOPS running on the ground~\cite{tsuda2020rendezvous}, which minimizes the fuel consumption taking into account the uncertainties in the asteroid properties and relative navigation for a 2.5$\sigma$ uncertainty. 

In our case, for the autonomous operation, we choose to apply a Lambert guidance~\cite{hawkins2012spacecraft}, which solves Lambert's problem after each measurement update using Izzo's algorithm~\cite{izzo2015revisiting}. Therefore, given the current state $\vec{r}(t_0)$, the desired final position $\vec{r}(t_f)$, and the time-to-go $t_{go}=t_f-t_0$, the Lambert problem can be solved to find the necessary velocity change at the current and the final states, $\Delta \vec{V}_0$ and $\Delta \vec{V}_f$, respectively. However, to manage the uncertainties, a Monte Carlo approach is applied considering the last estimate $\hat{\vec{X}}$ with its respective 3$\sigma$ uncertainty in $\vec{r}(t_0)$ and $\mu$. We consider 500 samples.

The commanded velocity changes will be taken from the mean value in each component of the vectors $\Delta \vec{V}_0$ and $\Delta \vec{V}_f$ from the Monte Carlo process. After each estimate update, a $\Delta \vec{V}_0$ will be performed, while the $\Delta \vec{V}_f$ will be executed a single time, scheduled after the last update before the orbital insertion. The obtained $\Delta \vec{V}$ can be accomplished by using the control command:
\begin{equation}
    \vec{u} = \frac{1}{\Delta t_u} \Delta\vec{V}.
\end{equation}
The time to initiate the $\Delta \vec{V}_f$ can be scheduled by considering a guaranteed maximum thrust level in each component, $u_m$, for the time $t_{\text{exec}}= t_f - \frac{ \Delta V_{\text{max}}}{u_m}$, where $\Delta V_{\text{max}}$ is the component with maximum magnitude of $\Delta \vec{V}_f$.

\subsection{Robust Path-following for Orbital Maintenance}

The second control law applied is a robust Keplerian path-following approach derived by Negri \& Prado \cite{negri2020novel}. As demonstrated in Negri \& Prado \cite{negri2021autonomous}, this control approach exhibits excellent performance in operating in close-proximity to small bodies, allowing the stabilization of otherwise unstable orbits even with a high level of uncertainties. Its robustness is derived from the application of sliding-mode control theory.

The sliding surface $\vec{s}$ is defined solely in terms of the integrals of motion of the two-body problem that are related to the geometry of the orbit. Thus, the true anomaly is treated as a free parameter, making the control a path-following approach, which is more suitable for orbital maintenance~\cite{negri2021autonomous}. Once the equilibrium condition for the sliding surface is reached ($\vec{s} = 0$), the controlled spacecraft asymptotically converges to the desired Keplerian geometry. The sliding surface is defined as~\cite{negri2020novel}:

\begin{equation}
\label{eqn:sliding_surface}
\vec{s} = \begin{bmatrix}
\tilde{\vec{e}} \cdot (\lambda_R \hat{r} + \hat{\theta}) \\
\tilde{h} \\
\hat{h}_d \cdot (\lambda_N \hat{r} + \hat{\theta})
\end{bmatrix}=0,
\end{equation}
where $\lambda_R>0$ and $\lambda_N>0$ determine the rate of convergence to the desired Keplerian geometry~\cite{negri2020novel}, $\tilde{\vec{e}} = \vec{e}-\vec{e}_d$ represents the error between, respectively, the current and desired eccentricity vectors, $\hat{h}_d$ is the desired specific angular momentum unit vector and $\tilde{h}=h-h_d$ is the error in the magnitude of the specific angular momentum. The unit vectors $\hat{r}$ and $\hat{\theta}$ are the unit vectors of the radial-transverse-normal (RTN) coordinates, $\hat{r}=\vec{r}/||\vec{r}||$ and $\hat{\theta}=\hat{h}\times \hat{r}$, in which the normal component $\hat{h}$ is the specific angular momentum unit vector.

The desired unit vector $\hat{h}_d$, defining the orbital plane, can be obtained as:
\begin{equation}
\label{eqn:ang_mom_ver}
\hat{h}_d = \begin{bmatrix}
\sin i_d \sin \Omega_d \\
- \sin i_d \cos \Omega_d \\
\cos i_d
\end{bmatrix},
\end{equation}
 by choosing a desired inclination $i_d$ and longitude of the ascending node $\Omega_d$. The current and desired angular momentum are obtained as $\vec{h}=|| \vec{r} \times \vec{v} ||$ and $h_d=\sqrt{\mu a_d (1-e_d^2)}$, respectively. Finally, the current and desired eccentricity vectors are respectively:
 \begin{equation}
\label{eq:evec}
\vec{e} = \frac{1}{\mu} (\vec{v} \times \vec{h} - \mu \hat{r}) = e \begin{bmatrix}
\cos \Omega \cos \omega - \sin \Omega \sin\omega\cos i \\
\sin \Omega \cos \omega + \cos \Omega \sin\omega\cos i \\
\sin\omega \sin i 
\end{bmatrix},
\end{equation}
for a desired $\Omega_d$, $\omega_d$ and $i_d$. 

In this study, the vectors $\hat{h}_d$ and $\vec{e}$ are defined differently depending on the asteroid being considered. For the asteroid Eros, these vectors are defined in the inertial frame, while for Bennu, they are defined in the orbit-fixed frame. This choice is made because the spacecraft's operation about Bennu is assumed to be in a sun-terminator orbit, while for Eros it is in a retrograde equatorial orbit. However, in practice, the difference between the two frames is minimal over the short time frame of the simulation, which spans only a few days. This control law is particularly useful because it can generate a Keplerian motion even in non-inertial frames. This has been demonstrated by Negri and Prado in their works \cite{negri2021autonomous,negri2020novel}, showcasing the effectiveness of the control law in different reference frames with little fuel expenditure for small-body missions.

Using the sliding surface in Equation \ref{eqn:sliding_surface}, robustness to bounded disturbances and asymptotic convergence to the geometry of a Keplerian orbit can be obtained using the control:
\begin{equation}
\label{eq:control_theo2}
\vec{u}  = - [RTN]^{-1} F^{-1} ( G + K \text{sgn}(\vec{s})) - \vec{f},
\end{equation}
$K \in \mathbb{R}^{3\times 3}$ is a diagonal positive definite matrix, the function $\text{sgn}(\vec{s}) \in \mathbb{R}^{3 \times 1}$ represents the sign function taken in each component of $\vec{s}$, the matrices $F$ and $G$ are defined by:

\begin{subequations}
\begin{align}
F &= \frac{1}{h\mu} \begin{bmatrix}
-h^2 & \left[2\lambda_R h-(\vec{v}\cdot\hat{r})r\right]h & -\mu r (\vec{e}_d\cdot\hat{h}) \\
0 & \mu rh & 0 \\
0 & 0& \mu r (\hat{h}_d\cdot\hat{h}) 
\end{bmatrix}, \\
G &=\frac{h}{r^2} \begin{bmatrix}
\tilde{\vec{e}}\cdot(\lambda_R\hat{\theta}-\hat{r}) -1 \\
0 \\
\hat{h}_d\cdot(\lambda_N\hat{\theta}-\hat{r})
\end{bmatrix},
\end{align}
\end{subequations}
and the matrix $[RTN]$ is a matrix that transforms from the Cartesian coordinates to RTN, defined as:

\begin{equation}
[RTN] = \begin{bmatrix}
\hat{r}^\text{T} \\ \hat{\theta}^\text{T} \\ \hat{h}^\text{T}
\end{bmatrix}
\end{equation}
with the superscript $\text{T}$ representing the tranpose.

The control in Equation \ref{eq:control_theo2} lies on the assumption that the magnitude of $\vec{h}$ and $\vec{r}$ are not zero. It also assumes that by defining an angle $\beta$ such that $\cos \beta = \hat{h} \cdot \hat{h}_d$, this angle is bounded by $\beta<90\degree$. If both assumptions hold, the matrix $F$ is always invertible~\cite{negri2020novel}. The diagonal gain matrix $K$ can be chosen to guarantee convergence for unknown bounded disturbances, assuming $\vec{d}$ written in RTN bounded such that $\lvert d_R \rvert < D_R$, $\lvert d_T \rvert < D_T$, and $\lvert d_N \rvert < D_N$, as \cite{negri2020novel}:

\begin{subequations}
\label{eqn:K_matrix}
\begin{align}
K_{1,1} &\geq  \frac{h}{\mu} D_R + \left\lvert \frac{2\lambda_R h-(\vec{v}\cdot\hat{r})r}{\mu} \right\rvert D_T + \frac{ r \left\lvert \vec{e}_d\cdot\hat{h}\right\rvert}{h}  D_N, \\
K_{2,2} &\geq r D_T, \\
K_{3,3} &\geq r \frac{\hat{h}_d\cdot\hat{h}}{h} D_N,
\end{align}
\end{subequations}
in which $K_{j,j}$, $j=1,2,3$, are the diagonal elements of the matrix $K$.

\subsection{Practical Considerations}
\label{sec:pract}

The main drawback of the sliding mode control is its discontinuous control input, which in many practical applications leads to chattering. That can be easily dealt with by allowing the system to converge within a boundary around the sliding surface, at the expanse of a bit of performance, as extensively documented in the literature \cite{slotine1991applied}. The most common approach is to substitute the sign function in Equation \ref{eq:control_theo2} with the saturation function:
\begin{equation}
\label{eq:sat}
\text{sat}(x;x^\star) = \begin{cases} 1, &\text{$x>x^\star$} \\\frac{x}{x^{\star}}, &\text{$-x^{\star}\leq x\leq x^\star$} \\ -1, &\text{$x<-x^\star$} \end{cases}.
\end{equation}
In the case of $\vec{s}$, we have a vector, so we define $\vec{x}$, $\vec{x}^\star \in \mathbb{R}^n$, and the equivalent of Equation \ref{eq:sat} for vectors is $\vec{g}=\text{sat}(\vec{x};\vec{x}^\star): \mathbb{R}^n \xrightarrow{} \mathbb{R}^n$, such that $g_i=\text{sat}(x_i;x_i^\star)$, i = 1,2,...$n$, for $i$ representing each component of the vector. In that way, the sign function in Equation \ref{eq:control_theo2} can be replaced by $\text{sat}(\vec{s};\vec{s}^\star)$.

As discussed by Negri \& Prado~\cite{negri2021autonomous}, it would be desirable that the orbit-keeping law accommodate turned-off periods, to save fuel and allow scientific measurements with no thruster interference. Thus, we choose to apply a hysteresis function inspired by the Schmitt trigger as a form of mathematical switcher that turns on/off the orbital maintenance control. We define that hysteresis function as:
\begin{equation}
\label{eq:hys}
\text{hys}(x) = \begin{cases} 1, & \text{$|x|>x^+$}  \\
0, & \text{$|x|<x^-$} \\
1, & \text{$x^-\leq |x| \leq x^+$ and hys($x_{p}$)$=1$}  \\
0, & \text{$x^-\leq |x| \leq x^+$ and hys($x_{p}$)$=0$}  
 \end{cases},
\end{equation}
where $\text{hys}(x_p)$ indicates the previous value of the hysteresis function. For a vector $\vec{x} \in \mathbb{R}^n$, we define $\vec{g}=\text{hys}(\vec{x} ): \mathbb{R}^n \xrightarrow{} \mathbb{R}^n$, such that $g_i=\text{hys}(x_i)$, i = 1,2,...$n$. Therefore, a practical form for the control law in Equation \ref{eq:control_theo2}, considering idle-thrusters periods, can be:

\begin{equation}
\label{eq:u_sat_hys}
\vec{u}_{RTN}  = - \left[ [RTN]^{-1} F^{-1} ( G + K \text{sat}(\vec{s};\vec{s}^\star) ) - \vec{f} \right] \lvert \lvert  \text{hys}\left(\vec{\chi}\right) \rvert \rvert_\infty,
\end{equation}
where $\text{hys}\left(\vec{\chi}\right)$ represents the hysteresis function taken in each component of a vector of orbital elements $\vec{\chi} = \begin{bmatrix} a & e & i & \omega & \Omega  \end{bmatrix}^\text{T}$. The operator $\lvert \lvert \cdot \rvert \rvert_\infty$ is simply the $\mathcal{L}_\infty$ norm, i.e.  $\lvert \lvert \vec{x} \rvert \rvert_\infty$ = $\max_i \lvert x_i \rvert$. This is just an ingenious mathematical way of saying that the control is turned off if all elements of $\text{hys}\left(\vec{\chi}\right)$ are zero, and turned on if any component of $\text{hys}\left(\vec{\chi}\right)$ is 1.

While it may be theoretically possible to use thrusters continuously for long periods, it is generally not recommended for several reasons. Firstly, continuous use of thrusters is unnecessary unless the spacecraft is using electric propulsion. Executing precise control commands is challenging due to inherent errors in the execution, which add up the longer the thrusters are active. This is particularly problematic when operating in dead reckoning mode with an autonomous spacecraft. Secondly, prolonged thruster activity can significantly affect the spacecraft's ability to accurately point its instruments (both for navigation and science) while they are on. Additionally, the plumes generated by the thrusters can interfere with the instruments. The advantage of the orbital maintenance control strategy we employ is its capability to effectively accommodate long periods of thruster idling, mitigating these issues, as shown in details in Negri \& Prado \cite{negri2021autonomous}, and reducing fuel expenditure.

An additional practical consideration is the spacecraft's inability to execute the calculated control commands precisely. For that sake, we assume a 1$\sigma$ dispersion of $3\%$ around each component of the calculated control commands. We also consider that the spacecraft has an operational envelope with control commands bounded within the magnitudes of 0.02 m/s$^2$ and 1$\times10^{-5}$ m/s$^2$ in each component. In the case a calculated control command component is below 1$\times10^{-5}$ m/s$^2$, no control is applied in that component. These chosen bounds are conservative assumptions, 0.02 m/s$^2$ is the same as a $20$ N thruster in a 1,000 kg spacecraft, equivalent to a single thruster of the twelve RCS (reaction control system) thrusters of Hayabusa 2 \cite{tsuda2013system}, or the six 22 N trajectory correction maneuver (TCM) thrusters of OSIRIS-REx \cite{bierhaus2018osiris}. The lower bound is equivalent to a 10 mN thruster for the same spacecraft mass, comparable to the four 10 mN thrusters equipped in Hayabusa 2. Given the conservative thrust bound of 0.02 m/s$^2$, which is equivalent to one out of twelve Hayabusa 2's thrusters, it is unlikely that the attitude of the spacecraft would have any impact on achieving or not achieving this thrust level. This conclusion is supported by the findings presented in the study of Negri et al. \cite{negri2022iac}.

\begin{table}[!h]
    \caption{Bounds of the control and uncertainty.}
    \label{tab:Bounds}
    \centering
    \renewcommand{\arraystretch}{1.5} 
    \begin{tabular}{l c}
    \hline
        \textbf{Variable} & \textbf{Assumed Bounds} \\
        \hline
        Control commands inaccuracy (1$\sigma$) & $3\%$ \\
        Control command magnitude bounds & 1$\times10^{-5}$ to 0.02 m/s$^2$ m/s$^2$ \\
        No control applied threshold & Below 1$\times10^{-5}$ m/s$^2$ \\
        Maximum thrust for orbital insertion & 0.45 m/s$^2$ \\
        \hline
    \end{tabular}
\end{table}

The only case when a dedicated action on the attitude's control system might be required is at the orbital insertion when we consider that the thrust level is allowed to be larger than the 0.02 m/s$^2$ bound for applying the $\Delta \vec{V}_f$ described in Section \ref{sec:lamb_guid}. In that case, it is considered that the spacecraft would use its main engine, and we consider a maximum thrust level of 0.45 m/s$^2$, which is equivalent to a 450 N engine in a 1,000 kg spacecraft. That thrust level could be delivered by three of the four 200 N main thrusters of OSIRIS-REx mission \cite{bierhaus2018osiris}. Although the orbital insertion would thus require a dedicated attitude action for targeting the main engine to $\Delta \vec{V}_f$, we consider it could be safely executed in the tens of minutes between a $\Delta \vec{V}_f$ calculation and its scheduled execution time. Therefore, we can still decouple the trajectory dynamics and control from the attitude~\cite{negri2022iac}. 

We also note that thruster selection is a matter of mission trade-off, and we make no other specific consideration beyond the described maximum and minimum thrust level bounds. In the case an ultra-precise orbit-keeping is needed, it is theoretically possible to employ colloidal micro-Newton thrusters, like the ones of LISA Pathfinder~\cite{racca2010lisa}. Although we will consider a continuous thrust signal in our simulations, this does not limit the use of chemical engines, as a PWPF (pulse-width pulse frequency) modulator~\cite{wie2008space} could translate the continuous thrust to discrete signs. Negri \& Prado \cite{negri2021autonomous} shows that the orbital maintenance control applied here, which can be argued as the most doubtful to operate in the discrete manner of a chemical engine, can efficiently operate with discrete signs using a PWPF modulator. The assumed values for the actual applied control are presented in Table \ref{tab:Bounds}.

The parameters employed in the orbital control law are presented in Table \ref{tab:Control}. It is noteworthy that the gain matrix $K$ ensures stability against perturbations bounded by 0.01 m/s$^2$ in any component, equivalent to the gravitational acceleration at 5.33 Ceres' radii from Ceres, the largest body in the asteroid belt. The upper bound in $\omega$ for the hysteresis is selected because we opt for circular orbits in the operations to be demonstrated later. The $360\degree$ bound implies that the argument of periapsis is allowed to vary freely.

In a real mission scenario, it is important to recognize that the hysteresis parameters can be adjusted as the spacecraft enhances its knowledge of the body and environmental properties. This adjustment can be integrated with a routine that computes stable orbits and adapts the hysteresis parameters to accommodate non-worrisome oscillations in the orbital elements. Although we do not delve into these considerations here to maintain focus on the scope of this work, such adjustments can be efficiently implemented based on studies in orbital mechanics about asteroids~\cite{scheeres2016orbital, kikuchi2021frozen}.

\begin{table}[!h]
    \caption{Parameters of the orbital maintenance control.}
    \label{tab:Control}
    \centering
    \renewcommand{\arraystretch}{2.0} 
    \begin{tabular}{c l}
    \hline
        \textbf{Variable} & \textbf{Assumed Value} \\
        \hline
        $K$ & \begin{tabular}{@{}c@{}}Equality in Eqs. \ref{eqn:K_matrix} for: \\ $[RTN]\vec{D}=\begin{bmatrix} 0.01 & 0.01 & 0.01 \end{bmatrix}^\text{T}$\end{tabular} \\
        $\vec{s}^\star$ & diagonal of K \\
        $\vec{\chi}^+$ & $\begin{bmatrix} 0.05 a_d & 0.1 & 7.0^\circ & 360.0^\circ & 7.0^\circ \end{bmatrix}$ \\
        $\vec{\chi}^-$ & $\begin{bmatrix} 0.01 a_d & 0.02 & 0.5^\circ & 0.5^\circ & 0.5^\circ \end{bmatrix}$ \\
        \hline
    \end{tabular}
\end{table}

\section{Analysis and Discussion}

This section analyzes and discusses the results obtained by applying the proposed autonomous GN\&C approach. In Section \ref{sec:far_approach} we make some comments on the far-approach phase of an asteroid mission, which is beyond the scope of this work. Section \ref{sec:close_approach_phase} shows results considering the close-approach, which are further analyzed in Section \ref{sec:monte_carlo} in Monte Carlo simulations. Finally, Section \ref{sec:shape_uncer} considers a different shape onboard to assess how the proposed autonomous GN\&C behaves.

\subsection{Some Comments on the Far-approach Phase}
\label{sec:far_approach}




By ``far-approach'', we consider the period of the mission when the spacecraft changes from heliocentric to relative navigation about the small-body, which is the same as phase 1 of the Hayabusa 2 campaign \cite{tsuda2020rendezvous}. That phase ends with the spacecraft at an arbitrarily far distance from the small body, when a preliminary assessment of the environment is made, such as identifying small moons, coarse asteroid shape reconstruction, asteroid's attitude determination, constraining the parameter of mass, and others. Depending on the asteroid's size and mass, the relative distance in that mission period should vary from a few million or hundred thousand kilometers up to thousands or hundreds of kilometers. From a guidance and control standpoint, this is the least critical phase of the mission. The dynamics should remain nearly constant throughout the operation, with the main forces being the sun's gravitational pull and the solar radiation pressure.

If the autonomy is restricted to the operation around the asteroid, that is when the transition from the ground to the autonomous operation takes place. In this case, the spacecraft would rely on the ground up to the moment when the asteroid is found as a point source in its optical cameras. After that, a hybrid approach could take place as the autonomous spacecraft could fast approach the body up to thousands or hundreds of kilometers with the supervision of the ground for a safe delegation of navigation responsibility. On the other hand, an onboard algorithm to search for the asteroid - considering its ephemeris uncertainty - should be used if the spacecraft has been autonomously operating since its deep-space cruise phase. 

The more or less constant dynamics in the far-approach phase make it easy to use simple guidance laws, such as an LQG or a ZEM/ZEV that considers the state's uncertainty, for approaching the body. As it will become apparent soon, with our future assumptions, there is no need for an exact orbit determination at this point. The critical aspect is to get close enough to the asteroid for the preliminary environment assessment. For current research describing embedded algorithms for making the asteroid's preliminary attitude, shape, and environment assessment, we refer the reader to other works \cite{dietrich2018robust,bercovici2019robust,baker2020limb,nesnas2021autonomous}.

Because cruise-phase deep space autonomous navigation is still an area of active research \cite{broschart2019kinematic,bhaskaran2020autonomous,di2021toward,andreis2022onboard}, and the arguably operational simplicity in terms of guidance and control for the far-approach phase, we abstain from making deep considerations on which profile the spacecraft adopts in that period and concentrate on the scope of this work - arguiably more challenging - that is when the spacecraft is fully autonomously navigating relative to the asteroid without ground supervision, similar to what was made in previous studies \cite{takahashi2021autonomous}.

\subsection{Close-approach Phase}
\label{sec:close_approach_phase}

A mission could have different profiles depending on the mission's goals and the availability of prior knowledge about the asteroid's environment and properties. We consider that after the preliminary environment assessment at the end of the far-approach phase, the spacecraft could opt between different profiles, depending on the general properties of the asteroid assessed at a distance. It is safe to assume that in the preliminary environment characterization, the spacecraft would be able to determine if the environment tends or not to be a solar radiation pressure-dominated one or if the body is elongated. We can then rely on consolidated results in the literature \cite{scheeres2012orbit,scheeres2014close,scheeres2016orbital} to propose that in the first case, the spacecraft would adopt a sun-terminator orbit, while in the latter, the choice would be for a retrograde equatorial orbit in the asteroid's inertial frame. Of course, this is a simplification to aid in the argument of this work. Other operational profiles could be embedded in the spacecraft, accounting for other types of systems (e.g., binaries) and mission goals.

We first consider a Bennu exploration case. The spacecraft departs from tens of kilometers from the asteroid, in a position $\begin{bmatrix} -150 & -150 & 10 \end{bmatrix}^\text{T}$ km with initial velocity $\begin{bmatrix} 0 & 0 & 0 \end{bmatrix}^\text{T}$ m/s, both in the orbit-fixed frame. The initial a priori value for the spacecraft's nominal position and velocity is set randomly at the beginning of the simulation for a $1\sigma$ dispersion in each component of $15$ km and $1$ m/s, respectively, around the real value. {We assume that no prior information about $\mu$ and $C_{SRP}$ is available, and their a priori values are arbitrarily set to 0.001 to stress the filter (i.e., no initial knowledge about them). Generally, the spacecraft will have a good initial estimate of $C_{SRP}$, but we want to stress the filter and the proposition.} The spacecraft is left in free fall until it accumulates four measurements, as shown in Fig. \ref{fig:state_machine}, considered the minimum batch size. After that, the autonomous GN\&C starts its operation, with the first batch of measurements increasing in size until it reaches the designed batch size of twenty measurements. Once the designed batch size is reached, it sequentially advances in time as new measurements arrive.

Following the first estimation, the Monte Carlo-Lambert guidance drives the spacecraft to orbital insertion by setting the final values in the position and velocity vectors corresponding to the ones of the desired orbit under the best estimate $\hat{\mu}$ of the gravitational parameter. The guidance's time-to-go is calculated for $t_f=48.5$ hours after the simulation starts. The chosen orbit is a 2 km circular sun-terminator orbit. The orbit-keeping control maintains the spacecraft in this orbit for up to $120.5$ hours ($\sim5$ days) from the initiation of the simulation. At this point, the orbit-keeping control modifies the desired orbital elements to correspond to those of a transfer ellipse. That transfer ellipse is calculated in the same manner explained in Negri \& Prado \cite{negri2021autonomous} for a Hohmann-like transfer. The desired orbital elements are changed once again when arbitrarily close to the transfer ellipse periapsis for the ones of an 800 m circular sun-terminator orbit. 

\begin{figure}[!h]
\centering
\subfloat[Real and nominal trajectories]{\includegraphics[width=.45\textwidth]{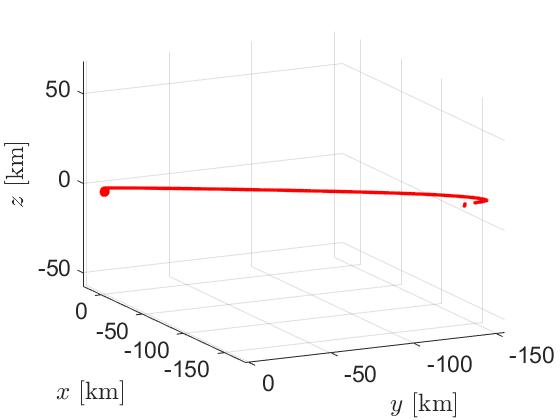}\label{Fig1a}} 
\subfloat[Real and nominal trajectories]{\includegraphics[width=.45\textwidth]{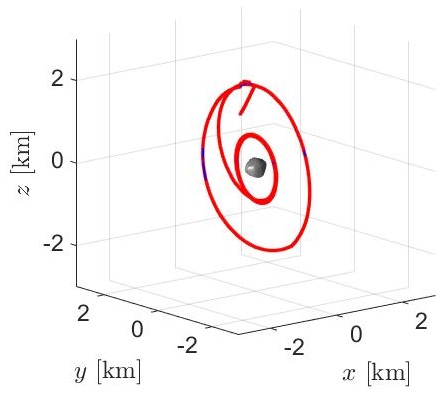}\label{Fig1b}}\\
\subfloat[Position estimation]{\includegraphics[width=.45\textwidth]{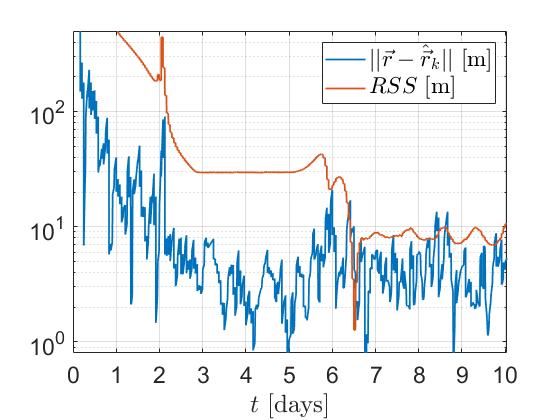}\label{Fig1c}} 
\subfloat[Velocity estimation]{\includegraphics[width=.45\textwidth]{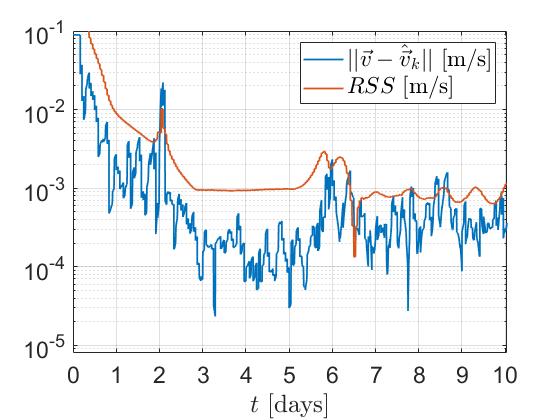}\label{Fig1d}}\\
\subfloat[$\mu$ estimation ratio]{\includegraphics[width=.33\textwidth]{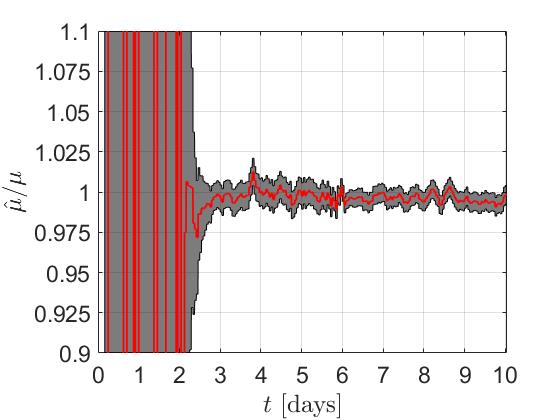}\label{Fig1e}} 
\subfloat[$C_{SRP}$ estimation ratio]{\includegraphics[width=.33\textwidth]{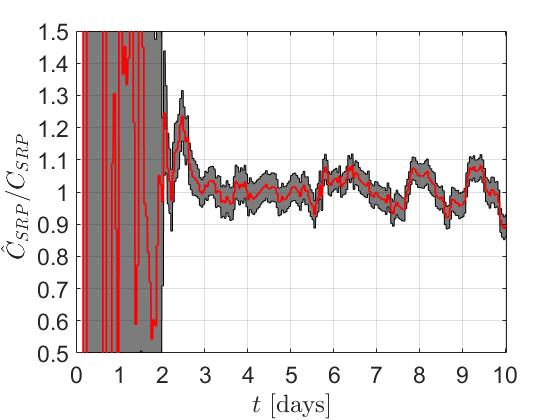}\label{Fig1f}}
\subfloat[Control commands]{\includegraphics[width=.33\textwidth]{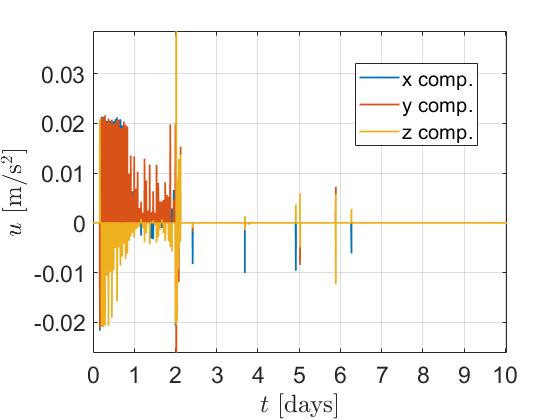}\label{Fig1g}}

\caption{Results for Bennu exploration.}
\label{Fig1}
\end{figure}

Figure \ref{Fig1} illustrates the simulation results for this scenario, considering a $1\sigma_R = 0.01$. In Figures \ref{Fig1a} and \ref{Fig1b}, where the actual trajectory is depicted in blue and the nominal trajectory in red, the successful operation is evident. The spacecraft is effectively inserted into orbit and subsequently undergoes a second orbit with a Hohmann-like transfer. Examination of Figures \ref{Fig1c} and \ref{Fig1d}, portraying the magnitude of errors in position and velocity represented by the blue line, reveals that these errors remain within a few meters and fractions of millimeters per second, respectively. Although the firing of the main engine introduces significant uncertainty in the spacecraft's state at $t=2$ days, this is swiftly resolved during the orbital operation.

{

In Figures \ref{Fig1c} and \ref{Fig1d}, the root sum of squares ($RSS$) of uncertainties in position and velocity is illustrated by orange lines. The $RSS$ is defined as the square root of the trace of the submatrices of the covariance matrix $P_k$ associated with either position (Fig. \ref{Fig1c}) or velocity (Fig. \ref{Fig1d}). Analyzing the behavior of the $RSS$ for position and velocity provides insights into the filter's performance. Overall, the filter demonstrates adequate performance, with uncertainties exhibiting magnitudes larger than the actual estimation errors.

However, an issue arises after the spacecraft's insertion into the second orbit, where the $RSS$ values drop below the estimation errors (between $t=6$ and 7 days). This suggests an overconfidence, which is later rectified as the $RSS$ magnitude returns to the same order as the actual errors. This outcome is expected in this drastic scenario, given that the batch estimator lacks process noise. At this closer proximity, the effects of unmodeled dynamics become more pronounced, particularly influenced by $\mathcal{J}_2$ and $\mathcal{C}_{22}$. Nevertheless, before the second orbit, the filter operates satisfactorily, successfully accommodating the uncertainties introduced by high thrust activities in the orbital insertion and transfer burns, as depicted in Fig. \ref{Fig1d}.

While we use the term ``overconfidence'' for simplicity in exposition, it is important to note, as argued in Section \ref{sec:estimation}, that the covariance matrix is not truly indicative of expected errors in the batch estimator when applied to nonlinear systems with unmodeled dynamics~\cite{schutz2004statistical}. Therefore, strictly speaking, it is not genuine ``overconfidence''. Additionally, the RSS and standard deviation presented in the figures are obtained from the $P_k$ matrix presented in Section \ref{sec:estimation}. We remind the reader that, because $P_k$ is only useful for conditioning the next estimation cycle, it is propagated and inflated to $P_{k+}$ as argued in the same section. Nevertheless, a very low level for the variances is still indicative of concern, as even with inflation, they can result in poor conditioning for the next estimation cycle.
}


Figures \ref{Fig1e} and \ref{Fig1f} illustrate the ratio of estimating the gravitational parameter and solar radiation pressure coefficient, respectively. The gray background represents the $3\sigma$ standard deviation, while the red line indicates the estimated value. It can be observed that the estimation of $\mu$ remains within approximately 2\%, and $C_{SRP}$ within 10\%, after a few measurements obtained in orbit. The significant variation in the estimation of $\mu$ and $C_{SRP}$ before orbital insertion is likely attributed to the relatively weak effects of gravitational and solar radiation pressure forces compared to the magnitude of the thrust applied to accelerate the spacecraft in a highly uncertain environment. It is important to note that the thrust is passed as $\Delta \vec{V}_j$s parameters to the filter, and they are not part of the estimates.

The path-following nature of the orbital station-keeping is very convenient for having idle-thrusters periods, allowing scientific measurements without interference from the propulsive system, as discussed in Negri \& Prado \cite{negri2021autonomous}. The spacecraft can experience hours of idle-thrusters as indicated by Figure \ref{Fig1g}, in which the components of the control commands are shown. The $\Delta V$ budget was 6.73 m/s, with most of it spent in the Monte Carlo-Lambert guidance, where the spacecraft still has a poor estimation performance, and in the orbital insertion main engine burn. This compares favorably with the $\approx$9-14 m/s reported by Takahashi \& Scheeres \cite{takahashiinproceedings} and the 22 m/s in the OSIRIS-REx mission for close-proximity operations~\cite{williams2018osiris}.

Suppose we extrapolate the $\approx$0.05 m/s spent by the spacecraft in the Hohmann-like transfer plus orbital maintenance in the 800 m orbit (tighter than the tightest 1 km orbit of OSIRIS-REx~\cite{williams2018osiris}). In that case, the spacecraft could still orbit Bennu, and make similar orbital transfers, for about 227 days before reaching the 9 m/s best scenarios of Takahashi \& Scheeres \cite{takahashiinproceedings}. The point here is not to advise the use of this paper's exact architecture and mission profile. Instead, it shows that a fully autonomous operation opens new possibilities for asteroid exploration. It is a paradigm shift in the current conservative approach of severely constraining uncertainties before close-proximity. 

Well-designed guidance and control laws can allow an autonomous spacecraft to have a bolder operation, even with a higher level of uncertainty in the navigation. On top of that, there is not a significant compromise in budget $\Delta V$ as one could expect. Therefore, a fully autonomous mission in close-proximity might not need a long 20 days preliminary survey phase like the OSIRIS-REx mission, and its 94 days approach phase could be potentially shortened~\cite{williams2018osiris}. {It is important to note that a real mission involves various additional requirements, beyond reducing uncertainties to a very low level, that impact the time expenditure during the preliminary survey and approach. However, from a GN\&C perspective, our study indicates that there is no indication that autonomous spacecraft studies should follow these same approach times to reduce uncertainties to a very low level.

 It is also crucial to emphasize that the comparison of these magnitudes with the OSIRIS-REx mission and other missions hereafter serves only to provide a notion of the order of magnitude of the $\Delta V$ budget in real mission cases. The intention is only to showcase that the architecture proposed in this study aligns well with the values expected within a similar kind of mission within the current paradigm. Of course, real missions have a lot more requirements, including very strict scientific requirements, that may impose a high burden in terms of the $\Delta V$ budget. }

\begin{figure}[!h]
\centering
\subfloat[Real and nominal trajectories]{\includegraphics[width=.45\textwidth]{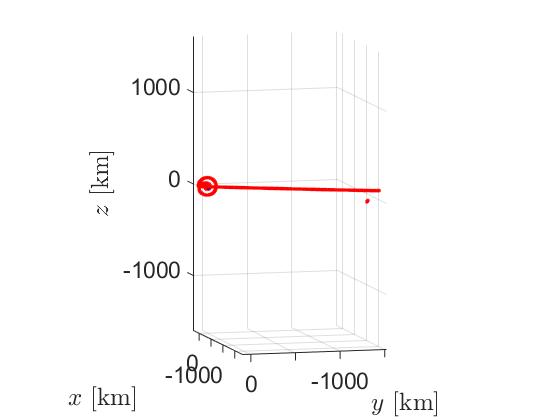}\label{Fig2a}} 
\subfloat[Real and nominal trajectories]{\includegraphics[width=.45\textwidth]{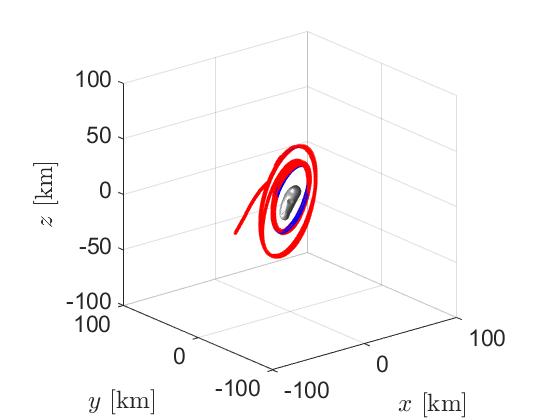}\label{Fig2b}}\\
\subfloat[Position estimation]{\includegraphics[width=.45\textwidth]{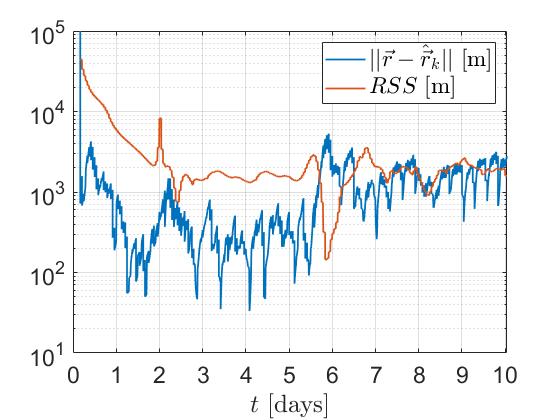}\label{Fig2c}} 
\subfloat[Velocity estimation]{\includegraphics[width=.45\textwidth]{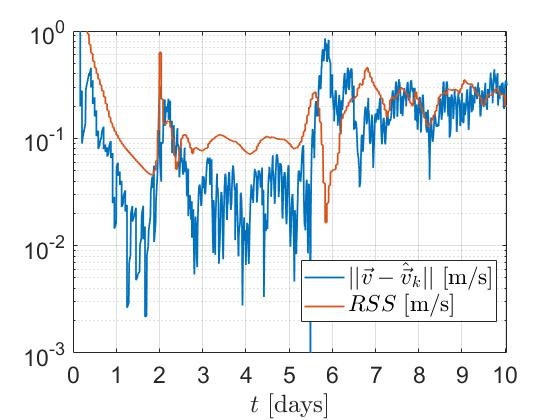}\label{Fig2d}}\\
\subfloat[$\mu$ estimation ratio]{\includegraphics[width=.45\textwidth]{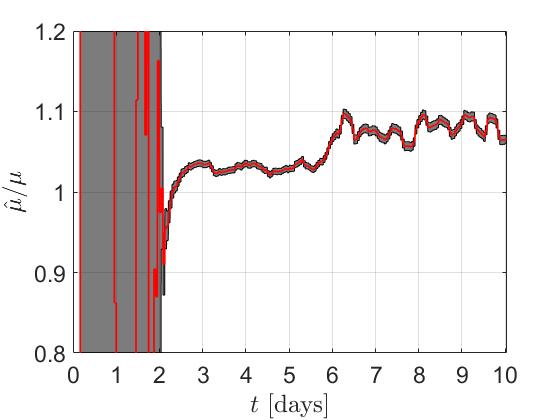}\label{Fig2e}} 
\subfloat[Control commands]{\includegraphics[width=.45\textwidth]{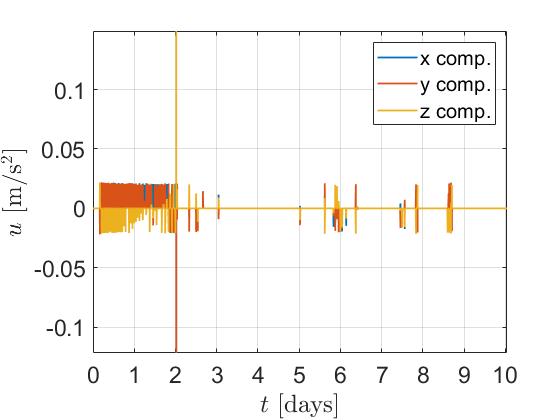}\label{Fig2g}}

\caption{Results for Eros exploration.}
\label{Fig2}
\end{figure}

\begin{figure}[!h]
\centering
\subfloat[Position estimation]{\includegraphics[width=.45\textwidth]{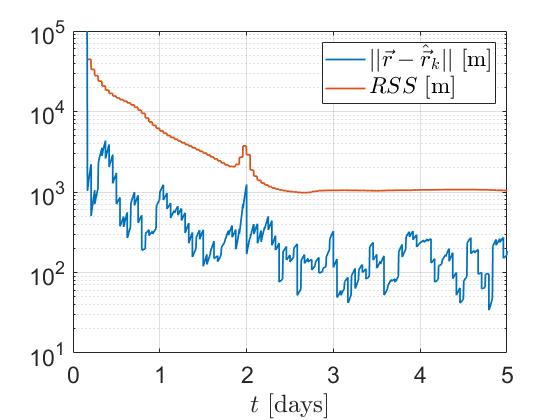}\label{Fig2addc}} 
\subfloat[Velocity estimation]{\includegraphics[width=.45\textwidth]{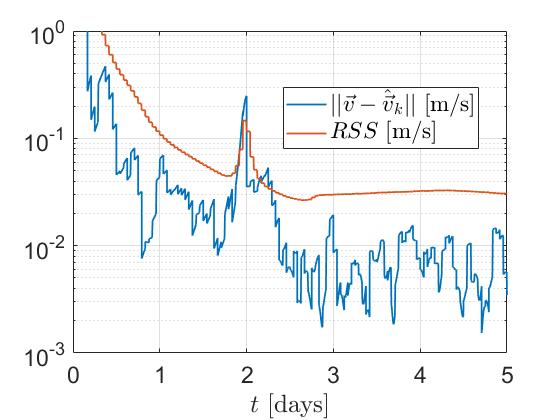}\label{Fig2addd}}\\
\subfloat[$\mu$ estimation ratio]{\includegraphics[width=.45\textwidth]{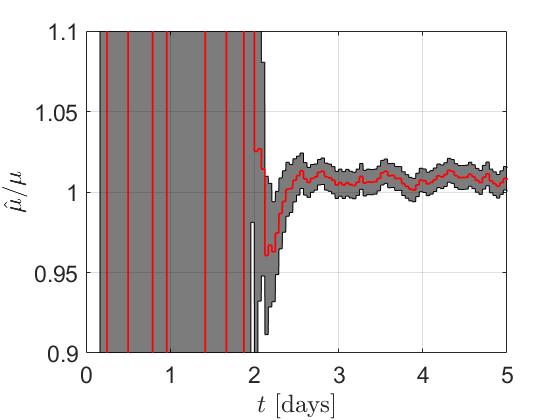}\label{Fig2adde}} 

\caption{Results for Eros exploration with a 100 km circular orbit.}
\label{Fig2add}
\end{figure}

{ 
We now shift our focus to a distinct target, the asteroid Eros, which stands in contrast to Bennu due to its elongated shape and significantly greater mass. Although some mission profile similarities with the Bennu scenario are maintained, the spacecraft now departs from $\begin{bmatrix} -1500 & -1500 & 100 \end{bmatrix}^\text{T}$ km. The a priori value for the position is randomly chosen with $1\sigma$ of 150 km in each component, while the velocity is chosen for a $1\sigma$ of 1 m/s. Orbital insertion occurs at 48.5 hours into an equatorial retrograde circular orbit of 50 km, considerably smaller than the initial eccentric orbit of over 300 km in the NEAR-Shoemaker mission~\footnote{\url{https://near.jhuapl.edu/NewMissionDesign/}}. Employing a Hohmann-like transfer, the spacecraft goes to a 30 km orbit, smaller than the NEAR-Shoemaker's closer 35 km orbit. As highlighted in Section \ref{sec:nominal}, the selection of Eros serves the dual purpose of not only examining the proposed architecture around an elongated body but also pushing its limits by capitalizing on Eros's high mass and the substantial influence of higher-order terms of the gravity field to assess the robustness of the proposal. The adoption of the two very tight orbits (50 and 30 km) underscores this specific objective.

Figure \ref{Fig2} presents the outcomes for the Eros scenario. The success of the operation is showcased in Figs. \ref{Fig2a} and \ref{Fig2b}. The position estimation error remains within a few hundred meters during the first orbit, as depicted in Figure \ref{Fig2c}, a level deemed adequate given the scale of the problem. However, a pronounced degradation in estimation accuracy is observed in the second orbit ($t>5$ days), reaching kilometer magnitudes. This deterioration is attributed to the omission of second-order terms of the gravity field ($\mathcal{J}_2$ and $\mathcal{C}_{22}$) in the nominal dynamics, particularly impactful at this proximity. 

If we assess the force due to the second-order gravity field on points along a sphere around Eros and calculate their mean magnitude, the contribution of the second-order gravity field can be said to be roughly 3\% of the main gravity term in the first orbit and 10\% in the second orbit. This is indeed a high level for the unmodeled dynamics, surpassing to many orders the cumulative effects of other forces~\cite{antreasian2016osiris} that are likely to not be included in the spacecraft's embedded equations of motion in a real application (albedo, infrared, attitude-dependent SRP, etc.).

The $RSS$ in Figs. \ref{Fig2c} and \ref{Fig2d} reflect again an overconfidence in estimates, while in the second orbit, due to the unmodeled dynamics. Nevertheless, the filter exhibits resilience and manages to cope with this challenging scenario, maintaining stability despite the compromised performance. It is crucial to stress that this analysis does not advocate for spacecraft operations in an environment with such expected levels of unmodeled dynamics in the filter's nominal equations. Real-world applications would involve inserting the spacecraft into a more distant and conservative orbit, with $\mathcal{J}_2$ and $\mathcal{C}_{22}$ as parameters to be estimated and considered into the nominal dynamics (as already discussed, their non-inclusion in this study is to assess the robustness of the proposal to the unmodeled dynamics). 

For example, Fig. \ref{Fig2add} illustrates a more realistic operational profile where the spacecraft is inserted into a 100 km orbit (NEAR-Shoemaker was inserted into a comparable orbit a month after rendezvous). In this case, the filter exhibits excellent performance, showcasing significant improvements in estimates. The errors in position, velocity, and gravitational parameter, are maintained around a hundred meters, a few millimeters per second, and 2\%, respectively. This scenario is more representative of what the spacecraft is likely to encounter. Including $\mathcal{J}_2$ and $\mathcal{C}_{22}$ in the nominal dynamics, the operation in orbits as tight as those in Fig. \ref{Fig2} could achieve performance levels similar to those depicted in Fig. \ref{Fig2add}.

The primary conclusion drawn from these results underscores the robustness of this architecture. If the spacecraft's autonomous system executive directs it to a challenging environment or a contingency arises, the filter remains non-divergent despite potential performance issues. The proposed guidance and control system can also effectively manage the situation. Other flight software algorithms may be developed to detect poor performance~\footnote{This is a great advantage of the batch estimator, as it allows for statistical analysis that can indicate a deterioration in the quality of estimation and measurement, as noted in Section \ref{sec:estimation}.}, prompting the spacecraft to transfer to a more conservative orbit or enter in safe mode. 

An illustrative example of a challenging environment is the spacecraft being erroneously directed to a tighter orbit than would be recommended~\footnote{It is improbable that the autonomous system's executive will adopt an approach tailored to each of the myriad potential situations a spacecraft may encounter in the challenging environment of small bodies. Most likely, it will be able to deal with a few main scenarios and adapt from them.}, such as during an operation around a body like comet 67P/Churyumov-Gerasimenko, where the added effects of third and higher-order terms of the gravity field and gas drag forces can have a pronounced impact in the dynamics. A contingency scenario could involve a fuel leakage, introducing a significant, uncertain, and undesirable magnitude in $\Delta V$. 

Perturbations in velocity, such as the example of fuel leakage, pose more significant challenges when relying solely on LiDAR and optical measurements. As there are no ground-based radiometric Doppler measurements, there is no direct measurement of velocity available. As evident in Figs. \ref{Fig1d}, \ref{Fig2d}, and \ref{Fig2addd}, issues of overconfidence in velocity estimation tend to be more pronounced. This challenge is exacerbated by using the measured maneuvers directly as parameters. If the $\Delta V_j$s are introduced into the estimates, it is expected to enhance the estimation, particularly in velocity performance. However, incorporating this substantial number of $\Delta V_j$s into the estimation process would significantly escalate the order of equations in the filter. This would involve inverting very large matrices, coupled with the complexity of handling a changing size of the estimates vector $\hat{\vec{X}}_k$. The size will vary from epoch to epoch, given the non-constant number of $\Delta V_j$s, adding intricacy to the system.

The estimation of $C_{SRP}$ for Eros is not presented in Fig. \ref{Fig2}, given that the solar radiation pressure ($SRP$) in this case is several orders of magnitude smaller than the higher-order terms of the gravity field. Regarding the gravitational parameter, the error remains below 5\% for the first orbit and 10\% in the second (see Figure \ref{Fig2e}). The significant variation in the estimated value of $\mu$ in the second orbit is attributed to $\mu$ acting as the acceleration bias used in JPL's AutoNav system~\cite{riedel2000autonomous}. Since this work does not implement an acceleration bias in the estimates, the filter compensates for the unknown dynamics by adding a bias to the estimated gravitational parameter. The same is seen in the first orbit to a lesser degree. This was expected to occur as Eros is an elongated body and the spacecraft is very close to it.


Future studies should investigate the performance of alternative filtering approaches~\footnote{For example, the one presented in Poore et al.~\cite{poore2004batch} or Markley and Carpenter~\cite{markley2012linear}.} that retain the advantages of the batch estimator, such as the demonstrated robustness, while potentially introducing process noise to enhance filter performance in challenging environments and contingencies.
}

{

Figure \ref{Fig2g} illustrates the control commands during the 10-day operation in close-proximity to Eros. Extended periods of idle thrusters are observed during the orbital phase of the mission. The $\Delta V$ budget is 76.77 m/s, with the Monte Carlo-Lambert guidance accounting for most of it. Despite operating with significant uncertainties about the spacecraft's state and environment, the 51.08 m/s spent by the Monte Carlo-Lambert guidance before the orbital insertion burn is reasonable. To provide context, NEAR-Shoemaker spent 50.38 m/s in the one year after the rendezvous burn (TCM-17), from TCM-18 to TCM-23~\footnote{\url{https://near.jhuapl.edu/NewMissionDesign/prpevent422.html}}. The 25.69 m/s $\Delta V$ in orbital insertion and maintenance is noteworthy. Compare this to NEAR-Shoemaker's 31.57 m/s for orbital insertion (OIM) and orbital operations (OCM-1 to OCM-25)~\footnote{\url{https://near.jhuapl.edu/NewMissionDesign/prpevent422.html}}. 

Once again, it is essential to recognize the differences between the scenarios. NEAR-Shoemaker had to adhere to numerous requirements, including scientific objectives, while our analysis did not face similar constraints. On the other hand, NEAR-Shoemaker had more precise radiometric data and a dedicated ground team planning each maneuver optimally, significantly reducing uncertainties and the $\Delta V$ budget. In summary, this comparison only provides a sense of the magnitude involved, indicating that this approach is not significantly more expensive than a standard mission scenario. Moreover, it is also crucial to note that this assessment aims to evaluate the robustness of the approach, and no systematic attempt has been made to find optimal parameters (control gains, covariance inflation, and others) for the setup, and we always opted for conservative assumptions (e.g., not including $\mathcal{J}_2$ and $\mathcal{C}_{22}$ terms in the filter, and large errors in executing the exact commanded control) following that intent.
}

\subsection{Monte Carlo Analysis}
\label{sec:monte_carlo}

To evaluate the robustness of the integrated autonomous GN\&C system, we conducted a Monte Carlo analysis with 500 samples. We introduced some modifications to the setup outlined in Section \ref{sec:close_approach_phase} to mitigate the processing time burden. The orbital insertion now takes place at $t=24.5$ hours, and the transition to the second orbit occurs at $t=48.5$ hours. The simulation concludes at 3 days. 

\begin{figure}[!h]
\centering
\subfloat[Real trajectories]{\includegraphics[width=.45\textwidth]{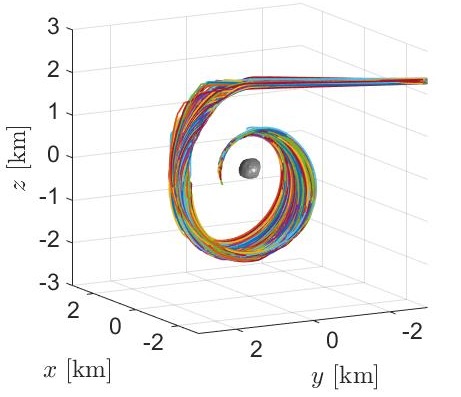}\label{Fig3a}} 
\subfloat[$\Delta V$ histogram]{\includegraphics[width=.45\textwidth]{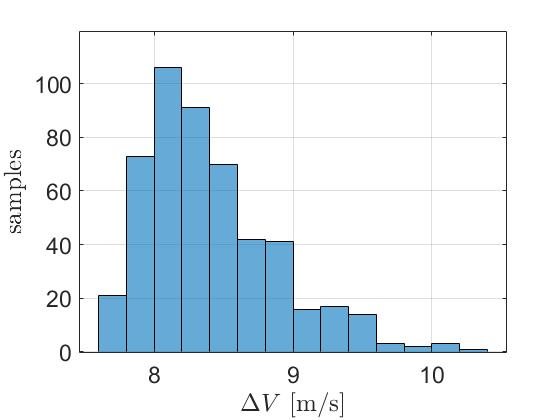}\label{Fig3b}}\\
\subfloat[Position estimation]{\includegraphics[width=.45\textwidth]{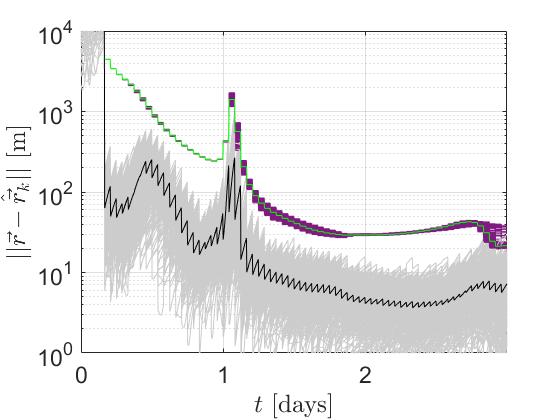}\label{Fig3c}} 
\subfloat[Velocity estimation]{\includegraphics[width=.45\textwidth]{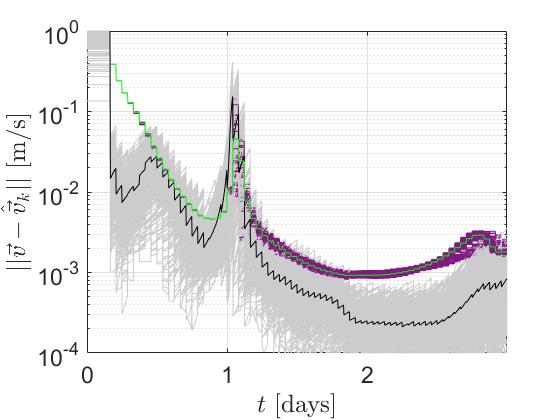}\label{Fig3d}}\\
\subfloat[$\mu$ estimation ratio]{\includegraphics[width=.45\textwidth]{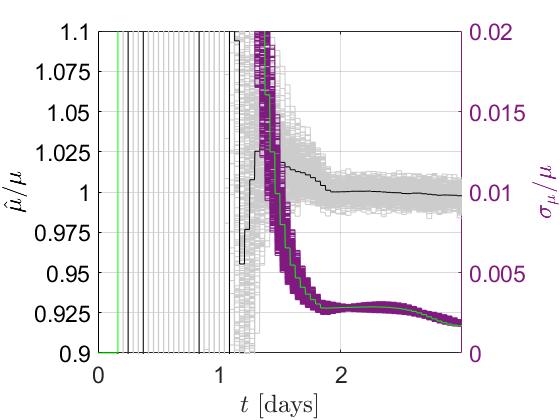}\label{Fig3e}} 
\subfloat[$C_{SRP}$ estimation ratio]{\includegraphics[width=.45\textwidth]{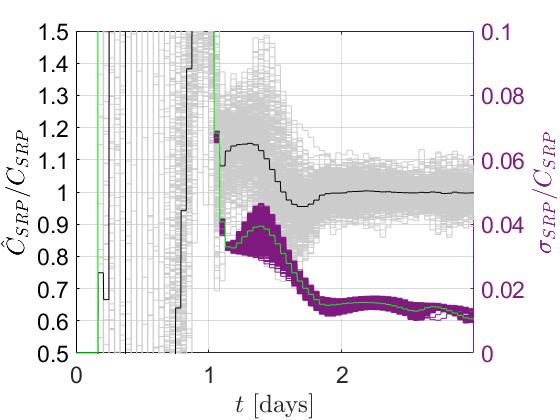}\label{Fig3f}}

\caption{Monte Carlo analysis for Bennu exploration.}
\label{Fig3}
\end{figure}

Figure \ref{Fig3} shows the results of the Monte Carlo simulation for Bennu. {In Figs. \ref{Fig3c} to \ref{Fig3f}, the estimation error of each sample is depicted in light gray, with the corresponding mean of all samples represented as a black line. Additionally, each sample of the position's RMS, velocity's RMS, and the $1\sigma$ standard deviation of $\mu$ and $C_{SRP}$ is shown in purple, with its respective mean in light green. For $\mu$ and $C_{SRP}$, the standard deviation is presented obeying a second y-axis to the right of plots \ref{Fig3e} and \ref{Fig3f}.

 All the samples are considered successful in accomplishing the proposition. Because the approaching time is half the one in Section \ref{sec:close_approach_phase}, the orbital injection error is higher. Thus, the autonomous spacecraft had initial trouble maintaining the orbit within reasonable bounds, which can be observed in the spikes of a few trajectories in Figure \ref{Fig3a}. However, once the spacecraft acquires a few measurements in a more favorable dynamical environment for estimation (i.e., with less control activity, closer proximity to the asteroid, and a more complex trajectory), it quickly stabilizes. The smooth transfer and maintenance in the tighter 800 m orbit become possible due to the improved accuracy and understanding gained from these measurements.

A histogram for the budget $\Delta V$ is shown in Figure \ref{Fig3b}. The consumption of $\Delta V$ is within 7.6 to 10.4 m/s having a mean of 8.41 m/s, with most of the consumption again in the Monte Carlo-Lambert guidance. Regarding the errors in estimating the spacecraft's state, they tend to remain in tens of meters and a few millimeters per second for the position and velocity, respectively, in the approach phase, as seen in Figs. \ref{Fig3c} and \ref{Fig3d} for the mean of the samples. 

The large orbital insertion burn introduces much uncertainty in the estimation, as expected from the analysis in Section \ref{sec:close_approach_phase}. The errors in position can spike to the order of a few hundred meters and centimeters per second for the velocity. Nevertheless, they rapidly decrease in the orbital operation, with the mean remaining a few meters and millimeters per second for the position and velocity, respectively. The means of the samples for estimating the constant parameters tend to be the real value of each of them, as can be checked in Figures \ref{Fig3d} and \ref{Fig3e}. 
}

\begin{figure}[!h]
\centering
\subfloat[Real trajectories]{\includegraphics[width=.45\textwidth]{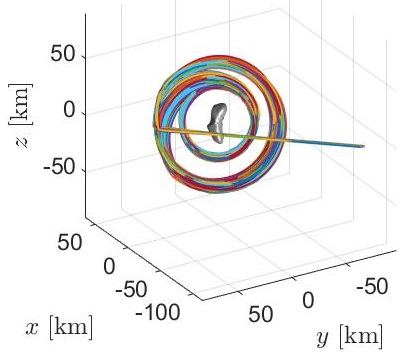}\label{Fig4a}} 
\subfloat[$\Delta V$ histogram]{\includegraphics[width=.45\textwidth]{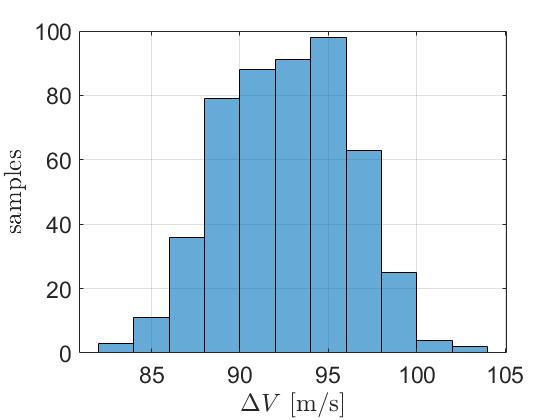}\label{Fig4b}}\\
\subfloat[Position estimation]{\includegraphics[width=.45\textwidth]{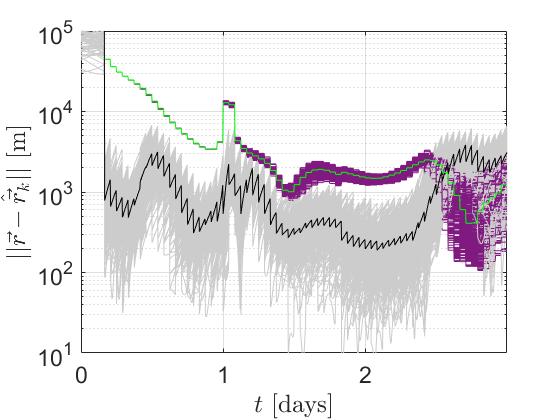}\label{Fig4c}} 
\subfloat[Velocity estimation]{\includegraphics[width=.45\textwidth]{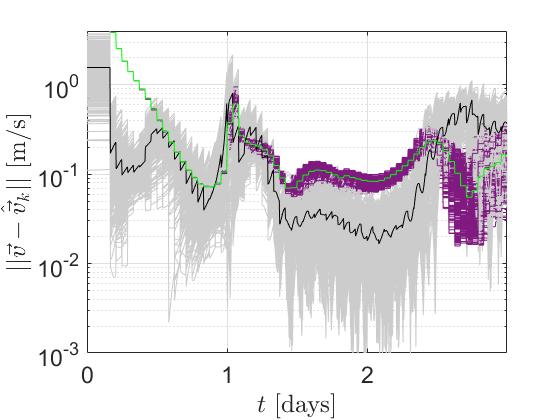}\label{Fig4d}}\\
\subfloat[$\mu$ estimation ratio]{\includegraphics[width=.45\textwidth]{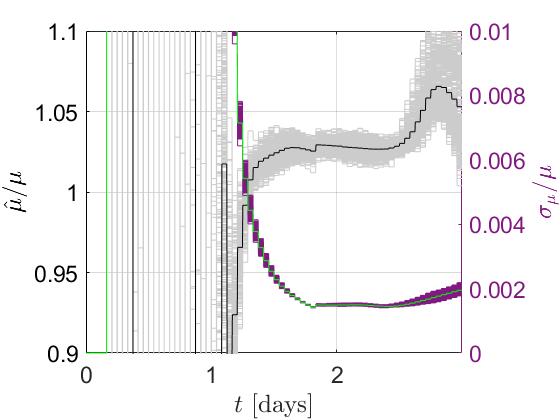}\label{Fig4e}} 

\caption{Monte Carlo analysis for Eros exploration.}
\label{Fig4}
\end{figure}

In the case of Eros, the results of the Monte Carlo analysis, as illustrated in Figure \ref{Fig4}, align with the discussion in Section \ref{sec:close_approach_phase}. Once again, the spacecraft executes its operation successfully. It is effectively inserted into orbit and completes the transfer smoothly, as depicted in Figure \ref{Fig4a}. The histogram in Figure \ref{Fig4b} reveals that the $\Delta V$ budget ranges between 82 and 104 m/s, with a mean of 92.66 m/s and a standard deviation of 3.59 m/s. In the first orbit, the mean error in estimation remains in the order of hundreds of meters and a few centimeters per second for position and velocity, respectively, as observed in Figures \ref{Fig4c} and \ref{Fig4d}. A degradation in estimation, with signs of overconfidence, is again noticeable for the second orbit due to the effects of the second degree and order of the gravity field neglected in the nominal dynamics. However, it is again crucial to emphasize that, for such an elongated asteroid in this tight orbit, assuming unmodeled dynamics of that magnitude is an over-conservative assumption. This assumption serves to demonstrate the robustness of the approach. The bias added to the estimation of $\mu$ due to $\mathcal{J}_2$ and $C_{22}$ can be noticeable in all samples in Figure \ref{Fig4e}.

\begin{figure}[!h]
\centering
\subfloat[Position estimation]{\includegraphics[width=.451\textwidth]{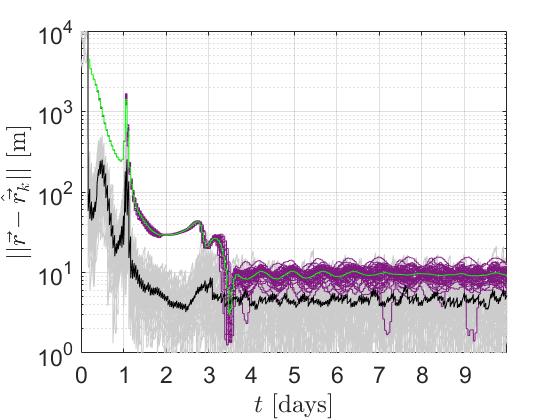}\label{Fig3addc}} 
\subfloat[Velocity estimation]{\includegraphics[width=.451\textwidth]{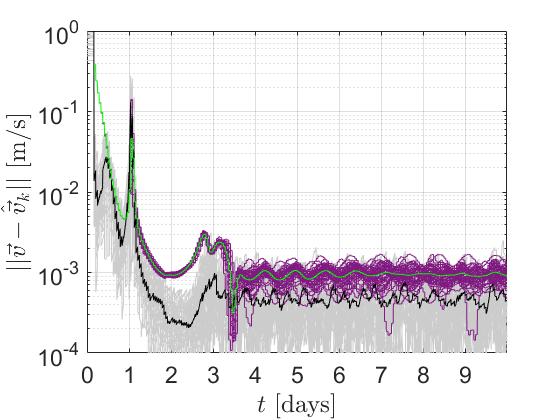}\label{Fig3addd}} \\
\subfloat[$\mu$ estimation ratio]{\includegraphics[width=.451\textwidth]{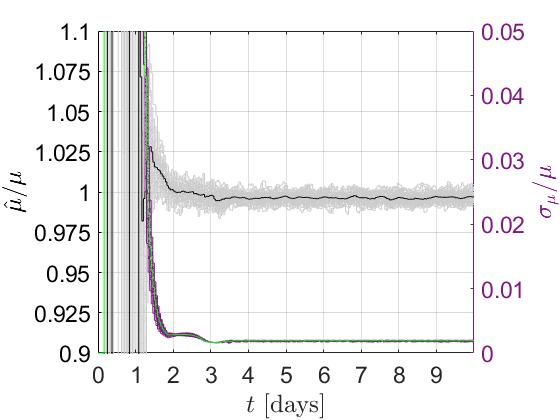}\label{Fig3adde}} 
\subfloat[$C_{SRP}$ estimation ratio]{\includegraphics[width=.451\textwidth]{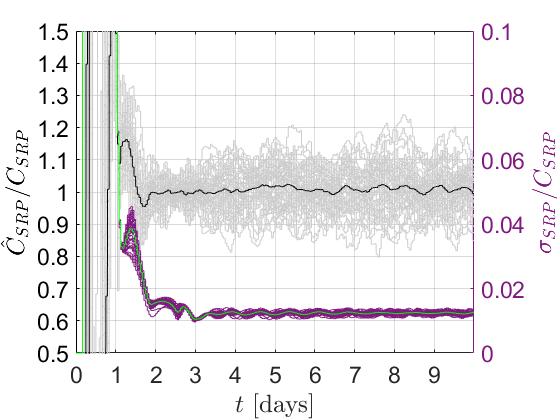}\label{Fig3addf}} 

\caption{Monte Carlo analysis for Bennu exploration, extending the operation time and considering 100 samples.}
\label{Fig3add}
\end{figure}

\begin{figure}[!h]
\centering
\subfloat[Position estimation]{\includegraphics[width=.451\textwidth]{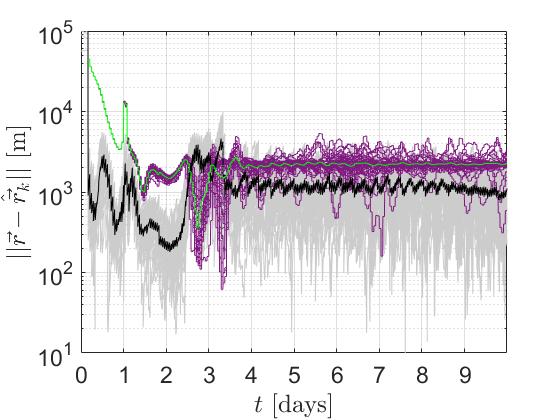}\label{Fig4addc}} 
\subfloat[Velocity estimation]{\includegraphics[width=.451\textwidth]{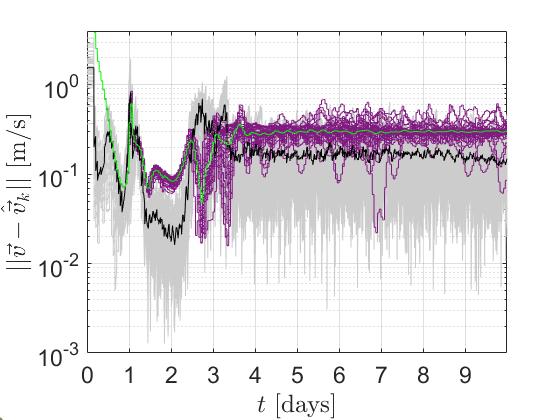}\label{Fig4addd}}\\
\subfloat[$\mu$ estimation ratio]{\includegraphics[width=.451\textwidth]{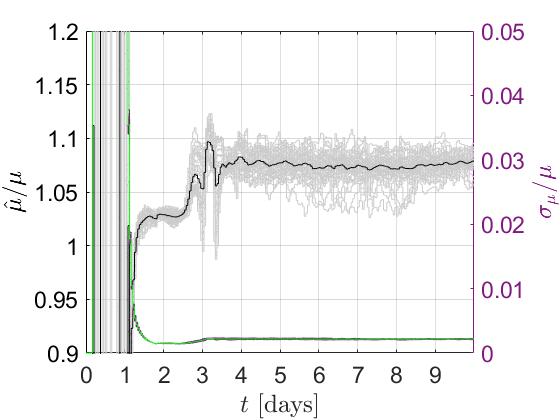}\label{Fig4adde}} 

\caption{Monte Carlo analysis for Eros exploration, extending the operation time and considering 100 samples.}
\label{Fig4add}
\end{figure}

\begin{figure}[!h]
\centering
\subfloat[Position estimation]{\includegraphics[width=.451\textwidth]{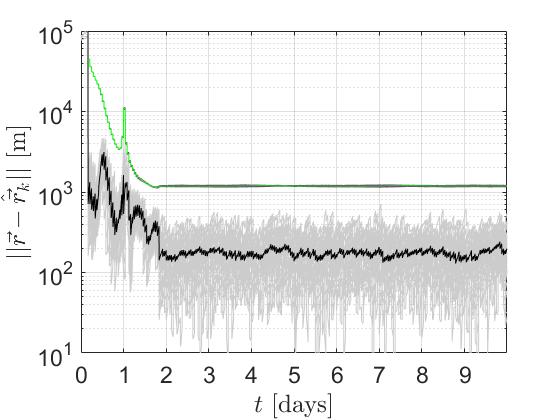}\label{Fig4addc2}} 
\subfloat[Velocity estimation]{\includegraphics[width=.451\textwidth]{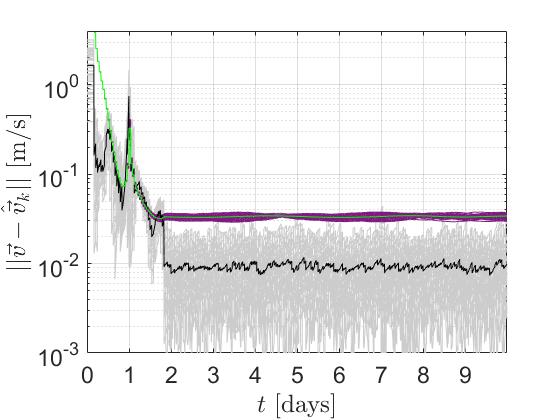}\label{Fig4addd2}}\\
\subfloat[$\mu$ estimation ratio]{\includegraphics[width=.451\textwidth]{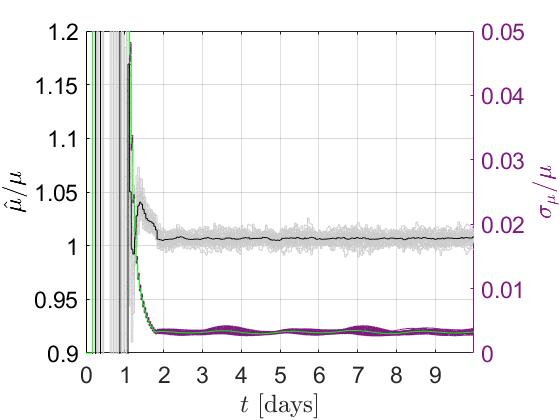}\label{Fig4adde2}} 

\caption{Monte Carlo analysis for Eros exploration, extending the operation time, considering 100 samples, and less drastic scenario.}
\label{Fig4add2}
\end{figure}

{

Due to the relatively short simulation time in the Monte Carlo runs, readers may question whether problematic behavior emerges after the initial 3 days. To address this, we extended the simulation to 10 days, considering a smaller set of 100 samples, to examine any signs of major issues beyond the 3-day mark. Figure \ref{Fig3add} illustrates the results for the Bennu exploration case, while Fig. \ref{Fig4add} displays the results for Eros. In the case of Bennu, an overconfidence issue is apparent in a few samples past the three-day mark, but subsequently, the RMS recovers, and estimation errors remain relatively constant within low margins. Similarly, for Eros, after overcoming the overconfidence issue around the two-day mark, the estimates stabilize. A few more cases of overconfidence happen in the Eros scenario when compared to Bennu, which was expected due to the elongated shape of Eros. This provides further evidence of the robustness of the proposed approach even with the conservative assumptions for the unmodeled dynamics level.

Once again, the selection of scenarios is intended to underscore the proposal's robustness under unfavorable conditions, while also incorporating conservative assumptions. These scenarios serve to demonstrate that, from a GN\&C perspective, an autonomous robotic spacecraft does not necessarily require extensive navigation campaigns to reduce uncertainties to very low levels. Moreover, even in challenging scenarios, the proposal exhibits resilience, indicating that a robotic spacecraft can handle those difficult conditions. Other algorithms could be developed to identify such challenging scenarios and guide the spacecraft toward more favorable operating conditions.

To illustrate this point, we conducted a Monte Carlo run with 100 samples for a more realistic scenario, the same depicted in Fig. \ref{Fig2add}. The spacecraft is inserted into a 100 km circular orbit around Eros. The results for this scenario are shown in Figure \ref{Fig4add2}. As evident, the filter performance is excellent, and the estimates are significantly more accurate. The reasons for this improvement were previously discussed in Section \ref{sec:close_approach_phase}.

It is crucial to emphasize that this is an initial assessment. While promising, additional challenges may surface as more complex simulations are conducted. These challenges could manifest when implementing such an approach in hardware for hardware-in-the-loop testing or when considering more intricate modeling for measurements in simulations. Although we took great care to be conservative in Section \ref{sec:measu}, it is possible that issues could emerge in more detailed models (e.g., incorporating bias). Nevertheless, this preliminary assessment is encouraging.
}

\subsection{Different Precision in Onboard Asteroid's Shape}
\label{sec:shape_uncer}

As discussed in Section \ref{sec:measu}, computationally light algorithms for shape reconstruction are being developed and investigated to be embedded as a substitution for the SPC technique used on the ground, which is computationally demanding and requires human intervention. For example, Panicucci \cite{panicucci2021autonomous} was able to reconstruct Bennu's shape using a silhouette-based algorithm considering a single Bennu rotation. The shape obtained had a mean error of $0.83\%$ and a maximum error of $4.49\%$~\cite[p. 79]{panicucci2021autonomous}, which is in line with the assumption of $\sigma_R=0.01$ made so far. In this case, Panicucci \cite{panicucci2021autonomous} assumes that the state and orientation of the spacecraft relative to the body are already solved. For the case the spacecraft's state needs to be determined altogether, Panicucci \cite[p. 125]{panicucci2021autonomous} applied a SLAM (simultaneous localization and mapping) algorithm and obtained a shape with a mean error of $3.21\%$ and maximum at $12.41\%$. Similar results were obtained by Nesnas et al.~\cite{nesnas2021autonomous} for a visual hull generated from a Shape-from-Silhouette algorithm.

\begin{figure}[!h]
\centering
\subfloat[Real trajectories]{\includegraphics[width=.45\textwidth]{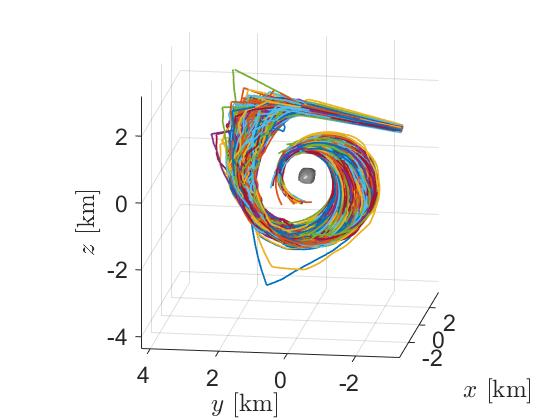}\label{Fig5a}} 
\subfloat[$\Delta V$ histogram]{\includegraphics[width=.45\textwidth]{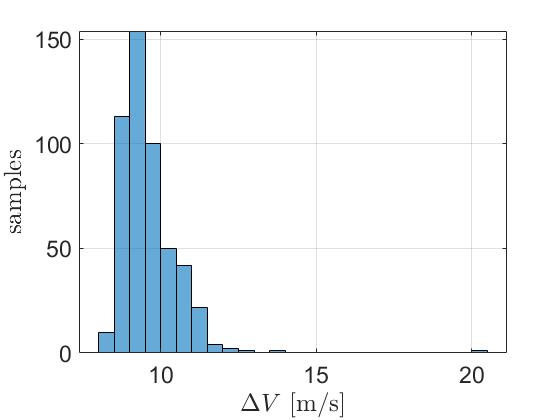}\label{Fig5b}}\\
\subfloat[Samples' statistics for position error estimation]{\includegraphics[width=.45\textwidth]{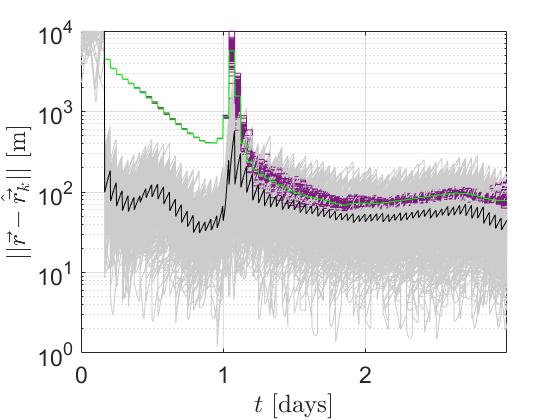}\label{Fig5c}} 
\subfloat[Samples' statistics for velocity error estimation]{\includegraphics[width=.45\textwidth]{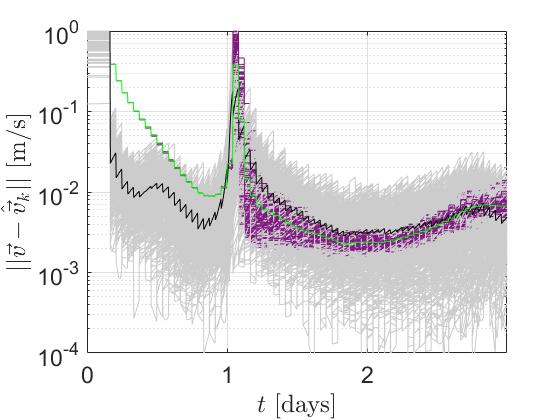}\label{Fig5d}}\\
\subfloat[Samples' statistics for $\mu$ estimation ratio]{\includegraphics[width=.45\textwidth]{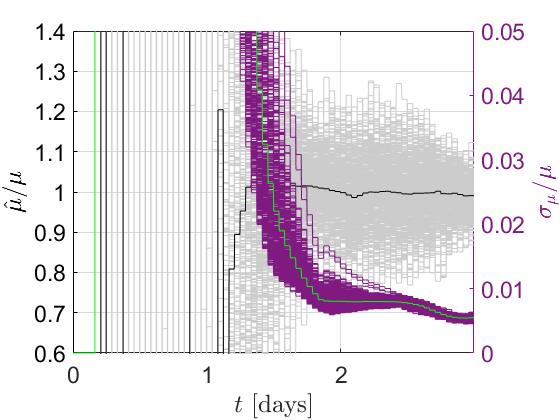}\label{Fig5e}} 
\subfloat[Samples' statistics for $C_{SRP}$ estimation ratio]{\includegraphics[width=.45\textwidth]{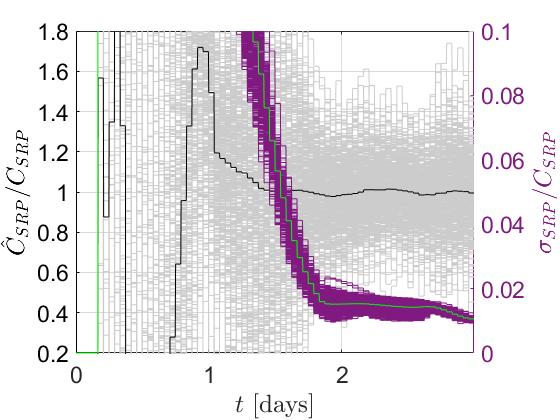}\label{Fig5g}}

\caption{Monte Carlo analysis for Bennu exploration considering $\sigma_R=0.05$.}
\label{Fig5}
\end{figure}

{

We conducted another Monte Carlo simulation for the asteroid Bennu, this time considering a $\sigma_R$ of 0.05 (i.e., 5\%) to evaluate the feasibility of using a shape model with considerable errors. Figure \ref{Fig5} presents the results for this scenario. It is noteworthy that all 500 samples were successful in the operation. However, upon closer inspection of Fig. \ref{Fig5a}, two significant observations can be made: orbital insertion errors tend to be more problematic, evidenced by substantial spikes in the trajectory after the orbital insertion maneuver; a few trajectories exhibit concerning behavior, displaying even larger spikes, some occurring hours after orbital insertion, resulting in the orbit-keeping control functioning more as guidance, directing the spacecraft to the desired orbit rather than effectively performing station-keeping. This degradation in performance is reflected in the $\Delta V$ histogram in Fig. \ref{Fig5b}, where the mean $\Delta V$ is 9.60 m/s. A few samples exhibit a $\Delta V$ larger than 12 m/s, with the highest reaching 20.32 m/s.

Figures \ref{Fig5c} and \ref{Fig5d} present estimation errors and RMS for position and velocity, demonstrating that the errors in estimation are generally a hundred meters and a centimeter per second. This represents a significant downgrade compared to Figs. \ref{Fig3c} and \ref{Fig3d}, where $\sigma_R$ was set to 0.01. The RMS values indicate overconfidence, suggesting that the covariance inflation for this scenario should be considerably larger. Regarding the parameters $\mu$ and $C_{SRP}$ in Figs. \ref{Fig5e} and \ref{Fig5g}, the estimation errors are now within 30\% and 80\%, respectively.

These results underscore the significance of accurate asteroid shape information in spacecraft operations around asteroids. While a rough shape, such as the 5\% range considered here, may still permit successful operations most of the time with minimal concern, it is not advisable. Even for an autonomous spacecraft, operating in close-proximity to an asteroid with imprecise shape information could be considered only in exceptional cases where promptly acquiring close-proximity data is imperative. In instances where the asteroid's shape falls within that order of magnitude, a more prudent approach would involve a slower approach to the target, minimizing the need for a large orbital insertion maneuver, and maintaining the spacecraft in a larger orbit, as in the 2 km first orbit of this example.

}

\section{Conclusions}

This work presents a paradigm shift in autonomous asteroid exploration studies by proposing a departure from the conventional approach of extensively reducing uncertainties before close-proximity operations. Instead, the focus is on leveraging autonomous robust guidance and control to handle uncertainties effectively. It is demonstrated that this alternative paradigm, centered around robust guidance and control, does not impose a significant fuel budget for small-body missions. Moreover, it is emphasized that the current state of technology is already capable of supporting these autonomous capabilities.

Aligned with recent studies aiming to develop robust algorithms for constraining an asteroid's shape within 1\% accuracy, our proposal exhibits great promise. Through Monte Carlo simulations, we evaluate the performance of the spacecraft in rapid approaching and transiting between tight orbits, starting with no prior knowledge about the asteroid's gravitational parameter. For the specific case of asteroid Bennu, the spacecraft successfully departs from a position hundreds of kilometers away, achieves orbital insertion into a 2000 m circular orbit within a day, and subsequently transits to an 800 m circular orbit within another day. Notably, the required $\Delta V$ remains remarkably low, ranging from 8 to 10 m/s, in comparison to 9-14 m/s reported in previous autonomous operation studies. Despite our conservative assumptions and the bold nature of the operation, the spacecraft achieves a few meters of positioning error and millimeters per second of velocity error in estimation. The estimation of the mass parameter remains below 2\%.

Extending the analysis to a larger and elongated asteroid, namely Eros, the spacecraft demonstrates the ability to depart from distances spanning thousands of kilometers and successfully execute insertion and transfers between tight orbits. The corresponding $\Delta V$ ranges from 80 to 105 m/s. Although the $\Delta V$ consumption for a larger body like Eros cannot be considered negligible compared to the small body case of Bennu, this performance is remarkable given the bold operation and conservative assumptions made during the assessment. The example of Eros also underscores the significance of accounting for at least the $\mathcal{J}_2$ and $\mathcal{C}_{22}$ terms of the gravity field for elongated asteroids, as large biases in estimation arise when the spacecraft approaches the target closely.

Furthermore, we evaluate the impact of a larger 5\% uncertainty in the asteroid's shape and observe that the proposed approach remains successful and robust. However, some abnormal trajectory instances are a source of concern. These findings emphasize that the primary bottleneck in this context remains the onboard shape reconstruction process.

Overall, this work contributes to the ongoing collective efforts aimed at demonstrating the feasibility of autonomous asteroid exploration, challenging the cautious and conservative approach inherited from current ground-in-the-loop missions. The results highlight the potential for rapid autonomous exploration within the existing technological landscape. Nonetheless, careful consideration must be given to the accuracy of the asteroid's shape and its impact on spacecraft operations, particularly during close-proximity maneuvers.

\section*{Acknowledgements}
The authors wish to express their appreciation for the support provided by grants $\#$ 2016/24561-0, 2017/20794-2 and 2021/10853-7 from S\~ao Paulo Research Foundation (FAPESP) and the financial support from the Coordination for the Improvement of Higher Education Personnel (CAPES).  

\bibliographystyle{plain}        
\bibliography{referencia}        

\begin{thebibliography}{10}

\bibitem{al2021validation}
MM~Al~Asad, LC~Philpott, CL~Johnson, OS~Barnouin, E~Palmer, JR~Weirich,
  MG~Daly, ME~Perry, R~Gaskell, EB~Bierhaus, et~al.
\newblock Validation of stereophotoclinometric shape models of asteroid
  (101955) bennu during the osiris-rex mission.
\newblock {\em The Planetary Science Journal}, 2(2):82, 2021.

\bibitem{andreis2022onboard}
Eleonora Andreis, Vittorio Franzese, and Francesco Topputo.
\newblock Onboard orbit determination for deep-space cubesats.
\newblock {\em Journal of guidance, control, and dynamics}, pages 1--14, 2022.

\bibitem{antreasian2016osiris}
PG~Antreasian, M~Moreau, C~Jackman, K~Williams, B~Page, and JM~Leonard.
\newblock Osiris-rex orbit determination covariance studies at bennu.
\newblock In {\em AAS Guidance and Control Conference}, pages 1--15, 2016.

\bibitem{archinal2018report}
BA~Archinal, CH~Acton, MF~A’hearn, A~Conrad, GJ~Consolmagno, T~Duxbury,
  D~Hestroffer, JL~Hilton, Randolph~L Kirk, SA~Klioner, et~al.
\newblock Report of the iau working group on cartographic coordinates and
  rotational elements: 2015.
\newblock {\em Celestial Mechanics and Dynamical Astronomy}, 130(3):1--46,
  2018.

\bibitem{baker2020limb}
DA~Baker and JW~McMahon.
\newblock Limb-based shape modeling and localization for autonomous navigation
  around small bodies.
\newblock In {\em 2020 Astrodynamic Specialist Conference, Lake Tahoe,
  AAS/AIAA}, 2020.

\bibitem{bercovici2019robust}
Benjamin Bercovici and Jay~W McMahon.
\newblock Robust autonomous small-body shape reconstruction and relative
  navigation using range images.
\newblock {\em Journal of Guidance, Control, and Dynamics}, 42(7):1473--1488,
  2019.

\bibitem{bhaskaran2012autonomous}
Shyam Bhaskaran.
\newblock Autonomous navigation for deep space missions.
\newblock In {\em SpaceOps 2012}, page 1267135. 2012.

\bibitem{bhaskaran2020autonomous}
Shyam Bhaskaran.
\newblock Autonomous optical-only navigation for deep space missions.
\newblock In {\em ASCEND 2020}, page 4139. 2020.

\bibitem{bhaskaran1998orbit}
Shyam Bhaskaran, S~Desai, P~Dumont, B~Kennedy, G~Null, W~Owen~Jr, J~Riedel,
  S~Synnott, and R~Werner.
\newblock Orbit determination performance evaluation of the deep space 1
  autonomous navigation system.
\newblock 1998.

\bibitem{bierhaus2018osiris}
EB~Bierhaus, BC~Clark, JW~Harris, KS~Payne, RD~Dubisher, DW~Wurts, RA~Hund,
  RM~Kuhns, TM~Linn, JL~Wood, et~al.
\newblock The osiris-rex spacecraft and the touch-and-go sample acquisition
  mechanism (tagsam).
\newblock {\em Space Science Reviews}, 214(7):1--46, 2018.

\bibitem{broschart2019kinematic}
Stephen~B Broschart, Nicholas Bradley, and Shyam Bhaskaran.
\newblock Kinematic approximation of position accuracy achieved using optical
  observations of distant asteroids.
\newblock {\em Journal of Spacecraft and Rockets}, 56(5):1383--1392, 2019.

\bibitem{cheng2018aida}
Andrew~F Cheng, Andrew~S Rivkin, Patrick Michel, Justin Atchison, Olivier
  Barnouin, Lance Benner, Nancy~L Chabot, Carolyn Ernst, Eugene~G Fahnestock,
  Michael Kueppers, et~al.
\newblock Aida dart asteroid deflection test: Planetary defense and science
  objectives.
\newblock {\em Planetary and Space Science}, 157:104--115, 2018.

\bibitem{board2019finding}
{Committee on Near Earth Object Observations in the Infrared and Visible
  Wavelengths}.
\newblock Finding hazardous asteroids using infrared and visible wavelength
  telescopes.
\newblock Technical report, 2019.

\bibitem{di2021toward}
Gianfranco Di~Domenico, Eleonora Andreis, Andrea~Carlo Morelli, Gianmario
  Merisio, Vittorio Franzese, Carmine Giordano, Alessandro Morselli, Paolo
  Panicucci, Fabio Ferrari, and Francesco Topputo.
\newblock Toward self-driving interplanetary cubesats: the erc-funded project
  extrema.
\newblock In {\em 72nd International Astronautical Congress (IAC 2021)}, pages
  1--11, 2021.

\bibitem{dietrich2018robust}
Ann~B Dietrich and Jay~W McMahon.
\newblock Robust orbit determination with flash lidar around small bodies.
\newblock {\em Journal of Guidance, Control, and Dynamics}, 41(10):2163--2184,
  2018.

\bibitem{dobrovolskis1996inertia}
Anthony~R Dobrovolskis.
\newblock Inertia of any polyhedron.
\newblock {\em Icarus}, 124(2):698--704, 1996.

\bibitem{feldhacker2017shape}
Juliana~D Feldhacker, Megan~Bruck Syal, Brandon~A Jones, Alireza Doostan, Jay
  McMahon, and Daniel~J Scheeres.
\newblock Shape dependence of the kinetic deflection of asteroids.
\newblock {\em Journal of Guidance, Control, and Dynamics}, 40(10):2417--2431,
  2017.

\bibitem{furfaro2015hovering}
Roberto Furfaro.
\newblock Hovering in asteroid dynamical environments using higher-order
  sliding control.
\newblock {\em Journal of Guidance, Control, and Dynamics}, 38(2):263--279,
  2015.

\bibitem{furfaro2013asteroid}
Roberto Furfaro, Dario Cersosimo, and Daniel~R Wibben.
\newblock Asteroid precision landing via multiple sliding surfaces guidance
  techniques.
\newblock {\em Journal of Guidance, Control, and Dynamics}, 36(4):1075--1092,
  2013.

\bibitem{gui2017control}
Haichao Gui and Anton HJ~de Ruiter.
\newblock Control of asteroid-hovering spacecraft with disturbance rejection
  using position-only measurements.
\newblock {\em Journal of Guidance, Control, and Dynamics}, 40(10):2401--2416,
  2017.

\bibitem{hawkins2012spacecraft}
Matt Hawkins, Yanning Guo, and Bong Wie.
\newblock Spacecraft guidance algorithms for asteroid intercept and rendezvous
  missions.
\newblock {\em International Journal of Aeronautical and Space Sciences},
  13(2):154--169, 2012.

\bibitem{izzo2015revisiting}
Dario Izzo.
\newblock Revisiting lambert’s problem.
\newblock {\em Celestial Mechanics and Dynamical Astronomy}, 121(1):1--15,
  2015.

\bibitem{kikuchi2021frozen}
Shota Kikuchi, Yusuke Oki, and Yuichi Tsuda.
\newblock Frozen orbits under radiation pressure and zonal gravity
  perturbations.
\newblock {\em Journal of Guidance, Control, and Dynamics}, pages 1--23, 2021.

\bibitem{leonard2019osiris}
Jason~M Leonard, Jeroen~L Geeraert, Brian~R Page, Andrew~S French, Peter~G
  Antreasian, Coralie~D Adam, Daniel~R Wibben, Michael~C Moreau, and Dante~S
  Lauretta.
\newblock Osiris-rex orbit determination performance during the navigation
  campaign.
\newblock In {\em 2019 AAS/AIAA Astrodynamics Specialist Conference}, pages
  1--20, 2019.

\bibitem{liounis2018limb}
Andrew~J Liounis.
\newblock Limb based optical navigation for irregular bodies.
\newblock In {\em 1st Annual RPI Workshop on Image-Based Modeling and
  Navigation for Space Applications, Troy, NY}, 2018.

\bibitem{markley2012linear}
F~Landis Markley and J~Russell Carpenter.
\newblock Linear covariance analysis and epoch state estimators.
\newblock {\em The Journal of the Astronautical Sciences}, 59:585--605, 2012.

\bibitem{mizuno2017development}
Takahide Mizuno, T~Kase, T~Shiina, Makoto Mita, Noriyuki Namiki, Hiroki Senshu,
  Ryuhei Yamada, Hirotomo Noda, Hiroo Kunimori, Naru Hirata, et~al.
\newblock Development of the laser altimeter (lidar) for hayabusa2.
\newblock {\em Space Science Reviews}, 208(1):33--47, 2017.

\bibitem{montenbruck2002satellite}
Oliver Montenbruck, Eberhard Gill, and Fh~Lutze.
\newblock Satellite orbits: models, methods, and applications.
\newblock {\em Appl. Mech. Rev.}, 55(2):B27--B28, 2002.

\bibitem{negri2022iac}
Rodolfo~B Negri, A~F B~A Prado, R~A~J Chagas, and R~V Moraes.
\newblock Six dof analysis for asteroid autonomous exploration.
\newblock In {\em 73rd International Astronautical Congress (IAC 2022)}, pages
  1--13, 2022.

\bibitem{negri2021autonomous}
Rodolfo~Batista Negri and Ant{\^o}nio~FBA Prado.
\newblock Autonomous and robust orbit-keeping for small-body missions.
\newblock {\em Journal of Guidance, Control, and Dynamics}, 45(3):587--598,
  2022.

\bibitem{negri2020novel}
Rodolfo~Batista Negri and Ant{\^o}nio Fernando Bertachini de~Almeida Prado.
\newblock Robust 3-d path following control for keplerian orbits.
\newblock {\em arXiv preprint arXiv:2012.01954}, 2020.

\bibitem{nesnas2021autonomous}
Issa~Antoine Nesnas, Benjamin~J Hockman, Saptarshi Bandyopadhyay, Benjamin~J
  Morrell, Daniel~P Lubey, Jacopo Villa, David~S Bayard, Alan Osmundson,
  Benjamin Jarvis, Michele Bersani, et~al.
\newblock Autonomous exploration of small bodies toward greater autonomy for
  deep space missions.
\newblock {\em Frontiers in Robotics and AI}, page 270, 2021.

\bibitem{panicucci2021autonomous}
Paolo Panicucci.
\newblock {\em Autonomous vision-based navigation and shape reconstruction of
  an unknown asteroid during approach phase}.
\newblock PhD thesis, Toulouse, ISAE, 2021.

\bibitem{poore2004batch}
Aubrey~B Poore, Benjamin~J Slocumb, Brian~J Suchomel, Fritz~H Obermeyer,
  Shawn~M Herman, and Sabino~M Gadaleta.
\newblock Batch maximum likelihood (ml) and maximum a posteriori (map)
  estimation with process noise for tracking applications.
\newblock In {\em Signal and Data Processing of Small Targets 2003}, volume
  5204, pages 188--199. SPIE, 2004.

\bibitem{racca2010lisa}
Giuseppe~D Racca and Paul~W McNamara.
\newblock The lisa pathfinder mission.
\newblock {\em Space science reviews}, 151(1-3):159--181, 2010.

\bibitem{riedel2000autonomous}
JE~Riedel, S~Bhaskaran, S~Desai, D~Hand, B~Kennedy, T~McElrath, and M~Ryne.
\newblock Autonomous optical navigation(autonav) ds 1 technology validation
  report.
\newblock {\em Deep Space 1 technology validation reports(A 01-26126 06-12),
  Pasadena, CA, Jet Propulsion Laboratory(JPL Publication 00-10), 2000}, 2000.

\bibitem{riedel2006autonav}
Joseph Riedel, Daniel Eldred, Brian Kennedy, Daniel Kubitscheck, Andrew
  Vaughan, Robert Werner, Shyam Bhaskaran, and Stephen Synnott.
\newblock Autonav mark3: Engineering the next generation of autonomous onboard
  navigation and guidance.
\newblock In {\em AIAA Guidance, Navigation, and Control Conference and
  Exhibit}, page 6708, 2006.

\bibitem{scheeres2014close}
Daniel~J Scheeres.
\newblock Close proximity dynamics and control about asteroids.
\newblock In {\em 2014 American Control Conference}, pages 1584--1598. IEEE,
  2014.

\bibitem{scheeres2016orbital}
Daniel~J Scheeres.
\newblock {\em Orbital motion in strongly perturbed environments: applications
  to asteroid, comet and planetary satellite orbiters}.
\newblock Springer, 2016.

\bibitem{scheeres2012orbit}
Daniel~Jay Scheeres.
\newblock Orbit mechanics about asteroids and comets.
\newblock {\em Journal of Guidance, Control, and Dynamics}, 35(3):987--997,
  2012.

\bibitem{scheeres2019autonomous}
DJ~Scheeres and JW~McMahon.
\newblock Autonomous architectures for small body exploration.
\newblock In {\em 2019 AAS/AIAA Astrodynamics Specialist Conference}, 2019.

\bibitem{schutz2004statistical}
Bob Schutz, Byron Tapley, and George~H Born.
\newblock {\em Statistical orbit determination}.
\newblock Elsevier, 2004.

\bibitem{slotine1991applied}
Jean-Jacques~E Slotine, Weiping Li, et~al.
\newblock {\em Applied nonlinear control}, volume 199.
\newblock Prentice hall Englewood Cliffs, NJ, 1991.

\bibitem{snyder2019electric}
John~Steven Snyder, Dan~M Goebel, Vernon Chaplin, Alejandro Lopez~Ortega,
  Ioannis~G Mikellides, Faraz Aghazadeh, Ian Johnson, Taylor Kerl, and Giovanni
  Lenguito.
\newblock Electric propulsion for the psyche mission.
\newblock 2019.

\bibitem{tajmar2004indium}
Martin Tajmar, A~Genovese, and W~Steiger.
\newblock Indium field emission electric propulsion microthruster experimental
  characterization.
\newblock {\em Journal of propulsion and power}, 20(2):211--218, 2004.

\bibitem{takahashiinproceedings}
Shota Takahashi and D.~Scheeres.
\newblock Autonomous proximity operations at small neas.
\newblock In {\em Proceedings of the 33rd International Symposium on Space
  Technology and Science}, 02 2022.

\bibitem{takahashi2021autonomous}
Shota Takahashi and Daniel~J Scheeres.
\newblock Autonomous exploration of a small near-earth asteroid.
\newblock {\em Journal of Guidance, Control, and Dynamics}, 44(4):701--718,
  2021.

\bibitem{takei2020hayabusa2}
Yuto Takei, Takanao Saiki, Yukio Yamamoto, Yuya Mimasu, Hiroshi Takeuchi,
  Hitoshi Ikeda, Naoko Ogawa, Fuyuto Terui, Go~Ono, Kent Yoshikawa, et~al.
\newblock Hayabusa2’s station-keeping operation in the proximity of the
  asteroid ryugu.
\newblock {\em Astrodynamics}, 4(4):349--375, 2020.

\bibitem{tsuda2020rendezvous}
Yuichi Tsuda, Hiroshi Takeuchi, Naoko Ogawa, Go~Ono, Shota Kikuchi, Yusuke Oki,
  Masateru Ishiguro, Daisuke Kuroda, Seitaro Urakawa, and Shin-ichiro Okumura.
\newblock Rendezvous to asteroid with highly uncertain ephemeris: Hayabusa2’s
  ryugu-approach operation result.
\newblock {\em Astrodynamics}, 4(2):137--147, 2020.

\bibitem{tsuda2013system}
Yuichi Tsuda, Makoto Yoshikawa, Masanao Abe, Hiroyuki Minamino, and Satoru
  Nakazawa.
\newblock System design of the hayabusa 2—asteroid sample return mission to
  1999 ju3.
\newblock {\em Acta Astronautica}, 91:356--362, 2013.

\bibitem{werner1997spherical}
Robert~A Werner.
\newblock Spherical harmonic coefficients for the potential of a
  constant-density polyhedron.
\newblock {\em Computers \& Geosciences}, 23(10):1071--1077, 1997.

\bibitem{werner1996exterior}
Robert~A Werner and Daniel~J Scheeres.
\newblock Exterior gravitation of a polyhedron derived and compared with
  harmonic and mascon gravitation representations of asteroid 4769 castalia.
\newblock {\em Celestial Mechanics and Dynamical Astronomy}, 65(3):313--344,
  1996.

\bibitem{wie2008space}
Bong Wie.
\newblock {\em Space vehicle dynamics and control}.
\newblock American Institute of Aeronautics and Astronautics, 2008.

\bibitem{williams2018osiris}
B~Williams, P~Antreasian, E~Carranza, C~Jackman, J~Leonard, D~Nelson, B~Page,
  D~Stanbridge, D~Wibben, K~Williams, et~al.
\newblock Osiris-rex flight dynamics and navigation design.
\newblock {\em Space Science Reviews}, 214(4):69, 2018.

\bibitem{zhang2019spacecraft}
Yonglong Zhang, Xiangyuan Zeng, and Fengdi Zhang.
\newblock Spacecraft hovering flight in a binary asteroid system by using fuzzy
  logic control.
\newblock {\em IEEE Transactions on Aerospace and Electronic Systems},
  55(6):3246--3258, 2019.

\end{thebibliography}



\end{document}